\documentclass[%
 reprint,
superscriptaddress,
amsmath,amssymb,
aps,
floatfix,
]{revtex4-2}

\usepackage[english]{babel}

\usepackage{amsmath,amssymb,amsthm}
\usepackage{tikz}

\usepackage{pgfplots}
\pgfplotsset{width=10cm,compat=1.9}

\usepackage{physics}
\usepackage{graphicx}
\usepackage[colorlinks=true, allcolors=blue]{hyperref}

\definecolor{blau}{RGB}{50,50,150}

\begin{document}
\preprint{APS/123-QED}

\title{Quantum simulation of one-dimensional fermionic systems with Ising Hamiltonians}

\author{Matthias Werner}
\email{matthias.werner@qilimanjaro.tech}
\affiliation{Qilimanjaro Quantum Tech., Carrer de Veneçuela, 74, Sant Martí, 08019 Barcelona, Spain}
\affiliation{Departament de Física Quàntica i Astrofísica (FQA), Universitat de Barcelona (UB), Carrer de Martí i Franqués, 1, 08028 Barcelona, Spain}
\affiliation{Institut de Ci\`encies del Cosmos, Universitat de Barcelona, ICCUB, Carrer de Martí i Franqu\`es, 1, 08028 Barcelona, Spain}

\author{Artur García-Sáez}
\affiliation{Qilimanjaro Quantum Tech., Carrer de Veneçuela, 74, Sant Martí, 08019 Barcelona, Spain}
\affiliation{Barcelona Supercomputing Center, Plaça Eusebi Güell, 1-3, 08034 Barcelona, Spain}

\author{Marta P. Estarellas}
\affiliation{Qilimanjaro Quantum Tech., Carrer de Veneçuela, 74, Sant Martí, 08019 Barcelona, Spain}

\thispagestyle{empty}
\pagestyle{empty}

\begin{abstract}
    In recent years, analog quantum simulators have reached unprecedented quality, both in qubit numbers and coherence times. Most of these simulators natively implement Ising-type Hamiltonians, which limits the class of models that can be simulated efficiently. We propose a method to overcome this limitation and simulate the time evolution of a large class of spinless fermionic systems in 1D using simple Ising-type Hamiltonians with local transverse fields. Our method is based on domain wall encoding, which is implemented via strong (anti-)ferromagnetic couplings $|J|$. We show that in the limit of strong $|J|$, the domain walls behave like spinless fermions in 1D. The Ising Hamiltonians are one-dimensional chains with nearest-neighbor and, optionally, next-nearest-neighbor interactions. As a proof-of-concept, we perform numerical simulations of various 1D-fermionic systems using domain wall evolution and accurately reproduce the systems' properties, such as topological edge states, Anderson localization, quantum chaotic time evolution and time reversal symmetry breaking via Floquet engineering. Our approach makes the simulation of a large class of fermionic many-body systems feasible on analogue quantum hardware that natively implements Ising-type Hamiltonians with transverse fields.

\end{abstract}

\maketitle

\section{Introduction}
In recent years, programmable quantum simulators have become capable of simulating quantum critical phenomena in many-body systems \cite{Bernien_2017, King_2018, Harris_2018}, including dynamical phase transitions \cite{Xu_2020}. While the devices can consist of thousands of qubits \cite{King_2022, King_2023}, these quantum simulations typically focus on Ising-type Hamiltonians with transverse fields, as these are native to quantum hardware platforms like superconducting flux qubits or neutral atoms. The simulation of 1D-systems of spinless fermions, or quantum spin chains, poses a challenge to these platforms due to a lack of non-stoquastic couplings, or limited control thereof.\\
Simulating fermionic systems is notoriously challenging even for quantum computers due to the overhead required to encode the non-local correlations. Several approaches have been developed to tackle this problem on the algorithmic level \cite{Abrams_1997, Ortiz_2001, Bravyi_2002, Verstraete_2005, Guseynov_2022} or by designing quantum computing hardware that automatically respects the Fermi-statistics \cite{González-Cuadra_2023}. The mentioned approaches mainly make use of the digital quantum computation paradigm, which typically requires both Trotterization \cite{Celeri_2023}, where the simulation time grows quadratically with the true evolution time \cite{Heyl_2019}, and non-stoquastic interactions, such as XX- and YY-terms in the Hamiltonians. The former point serves as a good motivation to consider continuous time, or analog, simulations of fermionic systems, while at the the same time the latter point makes fermionic Hamiltonians hard to implement on analog quantum hardware \cite{Ozfidan_2020, HitaPerez_2022}.\\
Here we present a continuous-time algorithm to simulate various one-dimensional systems of spinless fermions using only Ising-type Hamiltonians. While many fermionic systems found in nature are high-dimensional, various systems of fermions are known which display rich phase diagrams even in one dimension \cite{Mila_1993, Tsuchiizu_2002}. We show how Ising-type Hamiltonians can simulate the spectral properties, as well as the dynamical behaviour of a large class of one-dimensional fermionic systems. Our method allows to simulate spinless fermions in one dimension with fully programmable hopping energies, on-site disorder and nearest-neighbor interactions using only an Ising chain with nearest-neighbor- and next-nearest-neighbor couplings. This opens up the possibility to perform quantum simulations of disordered systems \cite{Barisic_2010, Vardhan_2017, Mierzejewski_2018, Lin_2018, Xu_2019, Kiefer_Emmanouilidis_2021} and systems that show topological phases \cite{Ganeshan_2013}. Beyond probing the properties of quantum many-body systems in a programmable device, our method can serve as a benchmark problem to quantify the quality of quantum hardware \cite{Chancellor_2022}. Furthermore, our approach allows to simulate the one-dimensional inhomogeneous Heisenberg chain, which is the workhorse model for quantum transport \cite{Sirker_2020, Kandel_2021, Mellak_2023}. This makes our method applicable as an integral part of larger engineered quantum systems.\\
The work is structured as follows: first, we will introduce the theory of how domain wall encoding \cite{Chancellor_2019} can be used to simulate spinless fermionic systems in one dimension to an arbitrarily close approximation. Then, we will numerically investigate the static properties of the approximation on the example of the Su-Schrieffer-Heeger (SSH) model \cite{Su_1979}, a key model for understanding topological excitations \cite{Meier_2016}. To investigate the simulation of dynamic properties, we will discuss simulations of the Aubry-Andre model \cite{Aubry_1980}, which has an extended-localized phase transition similar to the Anderson model. Furthermore, we investigate the influence of decoherence on the results by reproducing the dynamical phase transitions of the Aubry-Andre model. We test the method's ability to simulate interactions between particles in the inhomogeneous Heisenberg XXZ model and, lastly, we make use of Floquet engineering to realize exotic imaginary next-nearest-neighbor hopping terms.

\section{Method}
The algorithm is based on the realization, that domain wall encoding \cite{Chancellor_2019} can be used to simulate fermionic systems in one dimension. Domain wall encoding was initially conceived for quantum optimization, where it is used to encode discrete variables and where it out-performs various other encoding techniques \cite{Berwald_2022}. Since then, it has also been applied to solve field theories \cite{Abel_2021} and to model a single particle in a box \cite{Chancellor_2022}.

\subsection{Domain walls as non-interacting particles}
In domain wall encoding, the qubits are divided in two contiguous subsets, called domains, where in one domain all qubits are in the 0-state, while in the other all qubits are 1. The information is encoded in the location of the border of these domains, e.g. in a system of $N$ sites the state $|m\rangle$, $m=1...N$, is encoded in $N-1$ qubits as
\begin{equation}
    \begin{aligned}
        |m\rangle \equiv |\underbrace{00...0}_{(m-1)-\text{times}} 1...1\rangle  \ .
    \end{aligned}
    \label{eq:DWStates}
\end{equation}
The $N-1$ qubits are constrained by a strong ferromagnetic coupling $J$
\begin{equation}
    H_{z} = h_1 \sigma_1^z + h_{N-1} \sigma_{N-1}^z + J \sum_{n=1}^{N-2} \sigma_n^z \sigma_{n+1}^z
    \label{eq:Hz0}
\end{equation}
where the local fields on qubits $1$ and $N-1$ correspond to couplings to fixed virtual qubits of opposite orientation. In the conventional domain wall encoding by Chancellor \cite{Chancellor_2019}, the fields and couplings are chosen as $h_1 = -h_{N-1} = J$, such that the degenerate ground space of $H_z$ in Eq. \eqref{eq:Hz0} is spanned by the valid domain wall states Eq. \eqref{eq:DWStates}, where there is exactly one domain wall.\\
Many analog quantum computing platforms introduce dynamics to the diagonal target Hamiltonian via the transverse field Hamiltonian
\begin{equation}
    H_x = -t\sum_{n=1}^{N-1} \sigma_n^x \ .
    \label{eq:TransField}
\end{equation}
Consider the projector into the domain wall subspace
\begin{equation}
    P_{DW} = \sum_{m=1}^N |m\rangle \langle m | \ .
    \label{eq:ProjDW}
\end{equation}
Then it is easy to see that, in the case of a single domain wall, one finds
\begin{equation}
    P_{DW} H_x P_{DW} = -t\sum_{m=1}^{N-1} |m+1\rangle \langle m | + |m\rangle \langle m+1| \ .
\label{eq:ProjHxSingleDW}
\end{equation}
We find that for sufficiently strong ferromagnetic coupling, $|J| \gg |t|$, the $N$-dimensional subspace of the Hamiltonian $H_x + H_{z}$ corresponds to a single particle hopping on an finte chain of $N$ sites \cite{Chancellor_2022}. Here, the domain wall assumes the role of said particle and the transverse field Hamiltonian $H_x$ introduces the hopping. $H_x$ also couples the domain wall states to invalid qubit configurations, however these invalid states violate at least two additional couplings, and are thus energetically penalized.\\
While only the single domain wall subspace is energetically protected by virtue of being the ground space, the subspace of $M$ domain walls, also called $M$-excitation subspace, or $M$-subspace for shorthand, is separated by an energy of $\pm 4|J|$ from the $(M+2)$- and the $(M-2)$-excitation subspace respectively. This is because a bit flip either moves a domain wall by one site at no energy penalty, or it creates / destroys two domain walls, which raises or lowers the energy by $4 |J|$. This suppresses transitions between the $M$-subspaces in unitary evolution. In fact, below we will argue that the suppression of the transition probability is $O(|J|^{-4})$.\\
This means that the time-evolution, to a good approximation, will be contained in the $M$-excitation subspace, if the system is initialized in the $M$-subspace, allowing us to simulate multiple particles. Consider $P_{DW}^{(M)}$ the projector onto the subspace of $M$ excitations. Projecting $H_x$ into the $M$-excitation subspace, one finds that
\begin{equation}
    P_{DW}^{(M)} H_x P_{DW}^{(M)} = -t \sum_{m=1}^{N-1} \sigma_{n+1}^+ \sigma_n^- + \sigma_{n+1}^- \sigma_n^+
    \label{eq:ProjHxMultiDW}
\end{equation}
i.e. in the $M$-excitation subspace, $H_x$ acts identical to the hopping operator for spins. Applying the Jordan-Wigner transformation shows that in linear spin chains
\begin{equation}
    \sigma_{n+1}^+ \sigma_n^- + \sigma_{n+1}^- \sigma_n^+ = c_{n+1}^\dagger c_n + h.c.
\end{equation}
where $c_n^\dagger$, $c_n$ are the fermionic creation / annihilation operators on site $n$.\\
We can introduce on-site energies $\epsilon_n$ to the system via inhomogeneous couplings
\begin{equation}
    \begin{aligned}
        J_n &= J - \frac{1}{2} \epsilon_{n+1} \\
        h_1 &= J - \frac{1}{2} \epsilon_1 \\
        h_{N-1} &= -(J - \frac{1}{2} \epsilon_N)
    \end{aligned}
    \label{eq:FieldsCouplings}
\end{equation}
i.e. the Hamiltonian Eq. \eqref{eq:Hz0} is replaced by
\begin{equation}
    H_{z} = h_1 \sigma_1^z + h_{N-1} \sigma_{N-1}^z + \sum_{n=1}^{N-2} J_n \sigma_n^z \sigma_{n+1}^z \ .
\end{equation}
The local fields in Eq. \eqref{eq:FieldsCouplings} again correspond to couplings to fixed virtual spins. Consequently, the $M$-excitation subspace of the Hamiltoian
\begin{equation}
    \begin{aligned}
        H_{DW} = &-t \left( \sum_{n=1}^{N-1} \sigma_n^x  \right)+ h_1 \sigma_1^z \\ &+ h_{N-1} \sigma_{N-1}^z
        + \sum_{n=1}^{N-2} J_n \sigma_n^z \sigma_{n+1}^z
    \end{aligned}
    \label{eq:DomainWallHamiltonian}
\end{equation}
describes effectively $M$ fermions hopping on an linear chain according to the Hamiltonian
\begin{equation}
    H_{\text{Fermi}} = -t \sum_{n=1}^{N-1} \left( c_{n+1}^\dagger c_n + h.c. \right)  + \sum_{n=1}^N \epsilon_n c_n^\dagger c_n \ .
    \label{eq:HoppingHamiltonian}
\end{equation}
Note that the Ising Hamiltonian Eq. \eqref{eq:DomainWallHamiltonian} only requires a one-dimensional chain of spins with nearest-neighbor ZZ-coupling. Next we discuss how to create nearest-neighbor interactions between the domain walls / particles by introducing next-nearest-neighbor couplings between qubits.

\subsection{Domain walls as interacting particles}
An interaction between one particle on site $n$ and another particle on site $n+1$ can be induced by coupling the spins $n-1$ and $n+1$ in addition to the nearest-neighbor couplings. Consider the subsystem of the three spins $n-1$, $n$ and $n+1$ with those couplings. The particles / domain walls live on the bonds of qubits $n-1$, $n$ and $n$, $n+1$ respectively. If we choose both nearest-neighbor couplings to be equal to some $J_{NN}$, while the next-nearest-neighbor coupler is set to $J_{NNN}$, there are three spin configurations $000$, $001$ and $010$ with distinct energies, corresponding to no particle, one particle and two particles on the two sites under consideration. The other five configurations are equivalent by virtue of inversion and reflection symmetry. The respective energies need to satisfy the following system of equations
\begin{subequations}
    \begin{align}
        E(000) &= 2J_{NN} + J_{NNN} \overset{!}{=} 0\label{eq:E000}\\
        E(001) &= -J_{NNN} \overset{!}{=} 0 \label{eq:E001}\\
        E(010) &= -2J_{NN} + J_{NNN} \overset{!}{=} v\label{eq:E010}
    \end{align}
    \label{eq:NNEnergies}
\end{subequations}
in order to model some interaction strength $v$, which is unfortunately impossible, as this system of linear equations has no solution. However, if one relaxes Eq. \eqref{eq:E000} and Eq. \eqref{eq:E001} to $E(000)=E(001)$, one obtains the set of equations
\begin{subequations}
    \begin{align}
        2J_{NN} + J_{NNN} &\overset{!}{=} -J_{NNN} \\
        -2J_{NN} + J_{NNN} &\overset{!}{=} v
    \end{align}
\end{subequations}
which is solved by
\begin{subequations}
    \begin{align}
        J_{NNN} &= v/3 \\
        J_{NN} &= -v/3 \ .
    \end{align}
    \label{eq:NNCoulings}
\end{subequations}
The resulting energies are
\begin{subequations}
    \begin{align}
        E(000) = E(001) &= -\frac{v}{3} \\
        E(010) &= v \ .
    \end{align}
\end{subequations}
The energy penalty for having no or one excitation between the three spins results only in a constant energy shift of the full Hamiltonian, while the case of two excitations is $4v/3$ above that energy. Therefore, by replacing $v\rightarrow 3v/4$ and together with the couplings Eq. \eqref{eq:NNCoulings}, we can implement fermionic Hamiltonians with nearest-neighbor interactions of strength $v$ up to a constant energy shift $-v/4$.\\
Additionally, by choosing inhomogeneous transverse fields, i.e. by replacing $H_x$ from Eq. \eqref{eq:TransField} with
\begin{equation}
    H_x = -\sum_{n=1}^{N-1} t_n \sigma_n^x \ .
\end{equation}
we can model inhomogeneous hopping energies $t_n$. In summary, the Ising Hamiltonian
\begin{equation}
    \begin{aligned}
    &H_{DW} = -\sum_{n=1}^{N-1} t_n \sigma_n^x \\
    &+ \left( J - \frac{1}{2} \epsilon_1 - \frac{1}{4} v_1 \right) \sigma_1^z \\
    &- \left( J - \frac{1}{2} \epsilon_N - \frac{1}{4} v_{N-1} \right) \sigma_{N-1}^z \\
    &+ \frac{1}{4} v_1 \sigma_2^z - \frac{1}{4} v_{N-1} \sigma_{N-2}^z \\
    &+ \sum_{n=1}^{N-2} \left( J - \frac{1}{2} \epsilon_{n+1} - \frac{1}{4} v_n - \frac{1}{4} v_{n+1} \right) \sigma_n^z \sigma_{n+1}^z \\
    &+ \sum_{n=1}^{N-3} \frac{1}{4} v_n \sigma_n^z \sigma_{n+2}^z
    \end{aligned}
    \label{eq:DomainWallHamiltonianFull}
\end{equation}
with nearest-neighbor and next-nearest-neighbor interactions acting on $N-1$ spins simulates the one-dimensional fermionic Hamiltonian on $N$ sites
\begin{equation}
    \begin{aligned}
        &H_{\text{Fermi}} = -\sum_{n=1}^{N-1} t_n \left( c_{n+1}^\dagger c_n + h.c. \right) \\
        &+ \sum_{n=1}^N \epsilon_n c_n^\dagger c_n + \sum_{n=1}^{N-1} v_n c_n^\dagger c_n c_{n+1}^\dagger c_{n+1}
    \end{aligned}
    \label{eq:FermiHamiltonianFull}
\end{equation}
in the respective subspaces of $M$ domain walls for sufficiently large $|J|$, providing an efficient means to simulate fully programmable fermionic Hamiltonians of the type Eq. \eqref{eq:FermiHamiltonianFull}. Naturally, by choosing the model parameters as
\begin{equation}
    \begin{aligned}
        \epsilon_n &= 2 \begin{cases}
        \Tilde{\epsilon}_n - \Tilde{v}_n & n=1\\
        \Tilde{\epsilon}_{n} - \Tilde{v}_{n-1} & n = N \\
        \Tilde{\epsilon}_n - \Tilde{v}_n - \Tilde{v}_{n-1} & \text{else}
        \end{cases} \\
        v_n &= 4\Tilde{v}_n
    \end{aligned}
\end{equation}
and after applying the Jordan-Wigner transformations, the class of spin chains that can be simulated by the Ising chain $H_{DW}$ is
\begin{equation}
    \begin{aligned}
        H_{\text{spin}} = &-\sum_{n=1}^{N-1} t_n (\sigma_{n+1}^+ \sigma_{n}^- + \sigma_{n+1}^- \sigma_{n}^+) \\
        &+ \sum_{n=1}^{N-1} \Tilde{v}_n \sigma_{n}^z \sigma_{n+1}^z + \sum_{n=1}^N \Tilde{\epsilon}_n \sigma_n^z \ .
    \end{aligned}
    \label{eq:HSpinChain}
\end{equation}
Note that the Hamiltonian Eq. \eqref{eq:HSpinChain} can also be interpreted as a Hamiltonian of hard-core bosons, since excitations on spin chains follow the bosonic exchange statistics.\\
As a convention in this work, we will refer to one-dimensional Ising Hamiltonians such as the one in Eq. \eqref{eq:DomainWallHamiltonianFull} as Ising chains, while chains with energy transfer such as Eq. \eqref{eq:HSpinChain} are referred to as spin chains.\\
Note that the fields and couplings given in Eq. \eqref{eq:FieldsCouplings} only allow for an odd number of domain walls, however, an even number of excitations can be simulated by inverting the orientation of one of the virtual spins, which corresponds to changing the sign of the $\sigma_{N-1}^z$- and the $\sigma_{N-2}^z$-terms in Eq. \eqref{eq:DomainWallHamiltonianFull}.\\
Additionally, the excitation conservation argument also holds for one-dimensional spin chains with periodic boundary conditions, if the Ising chain is also periodic. In this case the number of particles is forced to be even, since on a periodic chain, the number of domain walls has to be even. This limitation, however, can be overcome by doubling the Ising chain by appending a copy of itself, before closing the periodic boundary. To simulate time evolution, the initial state has to be prepared in each subsystem respectively. This can be thought of as explicitly simulating two periods of the periodic model, instead of simulating one period explicitly and the infinite copies implicitly. The doubling of the particle number allows to simulate odd particle numbers in the original system with periodic boundaries. In case of periodic boundary conditions, some care has to be taken when preparing the initial state due to symmetry considerations. We provide more details on this question in appendix \ref{sec:AppendixA}.\\
Generally, it is possible to map all physically relevant observables that respect superselection rules to the domain wall picture, albeit the operators may not necessarily remain local. We discuss the construction method in appendix \ref{sec:AppendixB}.\\
While in this work we assume ferromagnetic couplings $J < 0$, note that our proposal also works for devices where strong anti-ferromagnetic couplings are available exclusively, or are favorable. If one exchanges the logical meaning of 0 and 1 on every other qubit and at the same time changes the sign of the nearest-neighbor couplers, the Hamiltonian Eq. \eqref{eq:DomainWallHamiltonianFull} can be effectively realized with $J > 0$. In case of periodic boundary conditions, this requires the number of qubits to be even, which, again, can be ensured by doubling the chain.\\
\begin{figure}
    \centering
    \includegraphics[scale=.22]{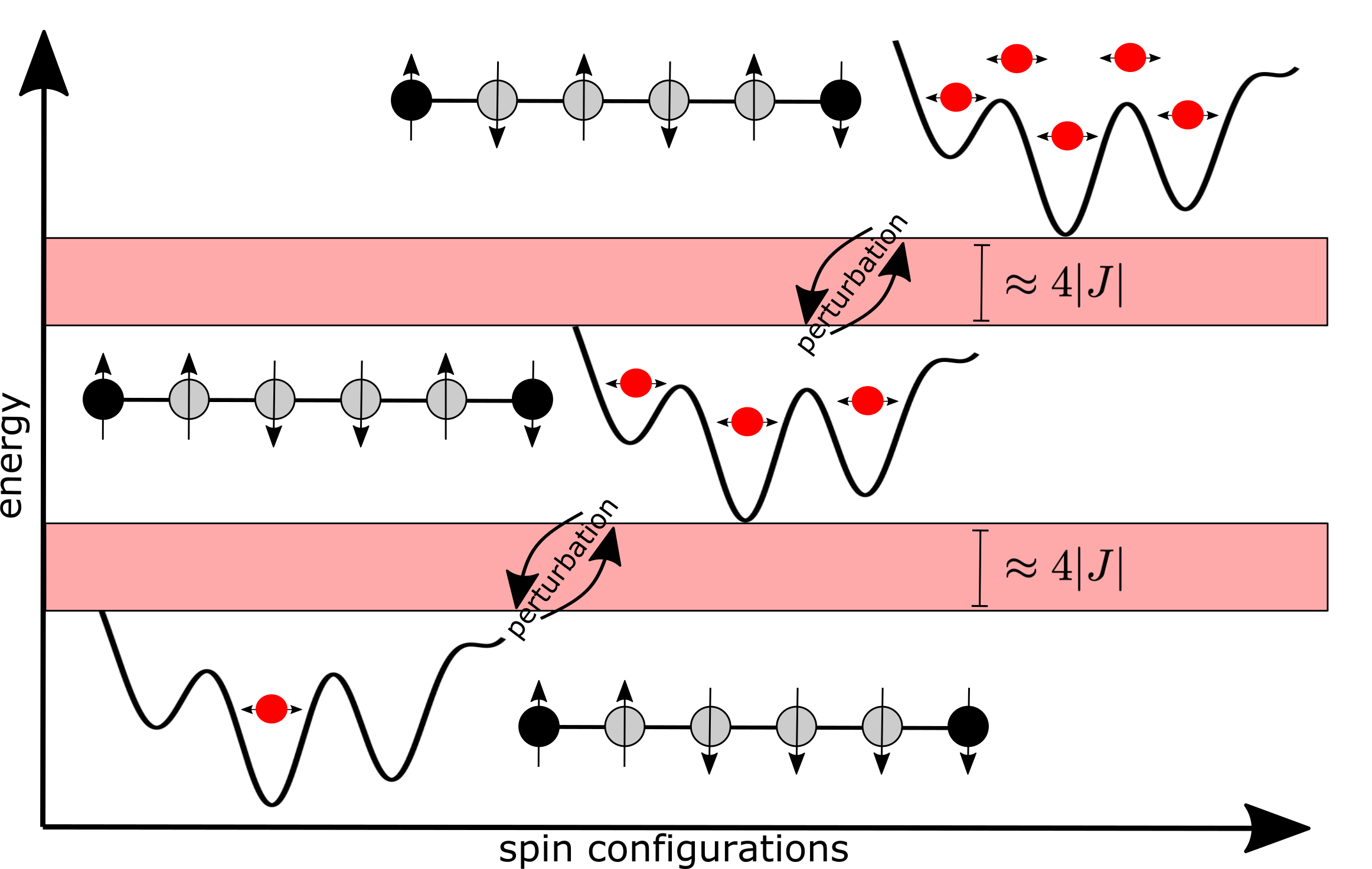}
    \caption{Illustration of the energy landscape of Hamiltonian $H_{DW}$ in Eq. \eqref{eq:DomainWallHamiltonian} and Eq. \eqref{eq:DomainWallHamiltonianFull}. As the number of domain walls increases, the energy landscape moves upwards by $4|J|$. For sufficiently large $|J|$, the subspaces corresponding to different numbers of domain walls are separated by energy gaps of $\approx 4|J|$ (red shaded areas). The transitions across these energy gaps are neglectable when considering time-evolution according to $H_{DW}$, however, the subspaces perturb each other and alter the time-evolution within each subspace. At a linear time-overhead, this perturbation can be made arbitrarily small.}
    \label{fig:EnergyLandscapeSketch}
\end{figure}
Figure \ref{fig:EnergyLandscapeSketch} illustrates the energy landscape of the model. As we will argue below based on perturbation theory and verify numerically, there are only neglectable transitions between the subspaces of different excitation number $M$ and instead the subspaces mainly act as perturbations on each other. The perturbation is due to undesired off-diagonal elements in $H_{DW}$ due to invalid spin-flips that create or destroy domain walls.\\
It should be highlighted that here we use our method to simulate systems with preserved particle number. However, it is possible to interpret the domain wall coupling $J$ as a chemical potential. The undesired off-diagonal elements due to individual spin flips create or destroy two domain walls on adjacent sites, which could be seen as $t_n (c_n^\dagger c_{n+1}^\dagger + c_n c_{n+1})$-terms in the Hamiltonian. As a consequence, our method could be used to simulate systems where the particle number parity is conserved, as long as the amplitudes of the particle hopping terms, $t_n (c_n^\dagger c_{n+1} + h.c.)$ , and the particle pair creation / annihilation terms, $t_n (c_n^\dagger c_{n+1}^\dagger + h.c.)$, are identical.

\section{Error analysis}
In the previous section we have proposed a method to simulate one-dimensional systems of spinless fermions using Ising Hamiltonians. This approach is an approximation, as the mapping between domain walls and fermions is only exact in the limit of $|J| \rightarrow \infty$. In this section, we will analyze the error of the approximation in more detail. To this end, we will employ the Schrieffer-Wolff transformation, which allows us to block-diagonalize the domain wall Hamiltonian $H_{DW}$ to first order in the undesired off-diagonal elements introduced by the transverse field. This allows us to understand the main source of error of the simulation and express the scaling of the error as a function of the simulation time $T$ and the domain wall coupling $|J|$. Based on these results, we show that the error of a quantum simulation can be arbitrarily reduced at the cost of moderate time-overhead.

\subsection{Perturbative analysis via Schrieffer-Wolff transformation}
\label{sec:ErrorAnalysis}
We will calculate an estimate of the fidelity of the simulation of Hamiltonian Eq. \eqref{eq:FermiHamiltonianFull}, when instead the system is evolved according to Hamiltonian Eq. \eqref{eq:DomainWallHamiltonianFull}. To this end, we will employ the Schrieffer-Wolff (SW) transformation \cite{Schrieffer_1966, Bravyi_2011}.\\
Since $H_{\text{Fermi}}$ in Eq. \eqref{eq:FermiHamiltonianFull} conserves the particle number, it is block-diagonal and we denote the projection of $H_{\text{Fermi}}$ into the $M$ particle subspace as $H_{\text{Fermi}}^{(M)}$. Similarly, we denote the projection of $H_{DW}$ Eq. \eqref{eq:DomainWallHamiltonianFull} into the subspace of $M$ domain walls as $H_{DW}^{(M)}$. $H_{\text{Fermi}}^{(M)}$ and $H_{DW}^{(M)}$ are equivalent up to constant energy shifts, i.e.
\begin{equation}
    \begin{aligned}
        H_{DW}^{(M)} = &H_{\text{Fermi}}^{(M)} \\
        &+ \left( (N-2M)J  - \frac{1}{4} \sum_{n=1}^{N-1} v_n \right) \mathbb{I}_M
    \end{aligned}
\end{equation}
where $\mathbb{I}_M$ is the identity matrix on the subspace of $M$ domain walls. The $J$-dependent term in the correction is due to the energy penalty from the domain wall coupling, while the sum over the $v_n$ is due to the energy offset from the nearest-neighbor interactions. Note that the energy correction due to the nearest-neighbor interactions is independent of $M$.\\
$H_{DW}$ additionally has block-off-diagonal elements due to spin flips that create or destroy two domain walls. We can denote these blocks as a perturbation $V_{M,M+2}$, i.e.
\begin{equation}
    \begin{aligned}
    H_{DW} &= \begin{pmatrix}
        H_{\text{DW}}^{(1)} & V_{1,3} &  &  \\
        V_{1,3} & H_{\text{DW}}^{(3)} & V_{3,5} &  \\
         & V_{3, 5} & H_{\text{DW}}^{(5)} & \ddots \\
         & & \ddots & \ddots &
    \end{pmatrix} \\
    &= H_{\text{Fermi}}^{\text{odd}} + D + V
    \end{aligned}
    \label{eq:DWHamiltonianDecomposed}
\end{equation}
where $V$ is the block-off-diagonal perturbation and $D$ a diagonal matrix that collects the energy offsets. $H_{\text{Fermi}}^{\text{odd}}$ denotes the fermionic Hamiltonian restricted to odd particle numbers. Note that here we discuss the setting of an odd number of domain walls, but the analysis extends readily to the setting of an even number of particles. Since the energy shift $D$ is constant in each block, we can ignore $D$ in the analysis of the dynamics within the respective subspaces. However, the energy shifts allow us to argue that the effect of the perturbation $V$ vanishes for large $|J|$. We can diagonalize the block-diagonal
\begin{equation}
    H_{\text{Fermi}}^{\text{odd}} + D = U \Lambda U^\dagger
\end{equation}
where $\Lambda$ is the diagonal matrix of eigenvalues and $U$ the unitary of eigenvectors. $U$ is also block-diagonal and we denote the projections into the respective subspaces as $U_M$. In this basis, the perturbation transforms into $\Tilde{V} = U^\dagger V U$, which remains block-off-diagonal where the entries are given by
\begin{equation}
    \Tilde{V}_{M,M+2} = U_M^\dagger V_{M,M+2} U_{M+2} \ .
\end{equation}
Note that the eigenbasis $U_M$ of $H_{\text{Fermi}}^{(M)}$ also diagonalizes $H_{DW}^{(M)}$ and the eigenvalues are the same up to the energy shift $(N-2M)J - \frac{1}{4} \sum_{n=1}^{N-1} v_n$.\\
In the SW transformation one aims at finding a unitary $\exp[S]$ such that $H_0 + H_1$ with a block-diagonal Hamiltonian $H_0$ under a block-off-diagonal perturbation $H_1$ becomes block-diagonal again. This allows to analyze the dynamics within a block by considering an effective Hamiltonian acting on the respective subspace. Consider the transformed Hamiltonian
\begin{equation}
	\Tilde{H} = \exp [S] (H_0 + H_1) \exp[-S] \ .
\end{equation}
The generator $S$ is to be determined such that $\Tilde{H}$ is block-diagonal. While it is common to consider the blocks of $\Tilde{H}$ to be the effective Hamiltonians, we will use a slightly different notion. As per convention, we will project $\Tilde{H}$ into the $M$-subspace, but additionally we will subtract the energy offsets $D$, i.e.
\begin{equation}
	H_{\text{eff}}^{(M)} := P_{DW}^{(M)} (\Tilde{H} - D) P_{DW}^{(M)} \ .
	\label{eq:DefEffHamiltonian}
\end{equation}
Since $D$ is constant within each $M$-subspace, this is equivalent to subtracting a constant from all eigenvalues and does not change the dynamics within the $M$-subspace described by the effective Hamiltonian. It does, however, simplify comparing the spectra of the effective and fermionic Hamiltonian.\\
Typically, the transformation $\exp[S]$ is hard to determine exactly and one settles for block-diagonalizing $H_0 + H_1$ up to some order of $H_1$. The transformed Hamiltonian is expanded
\begin{equation}
    \begin{aligned}
        \Tilde{H} &= \exp[S] (H_0 + H_1) \exp[-S] \\
        &= H_0 + H_1 + [S, H_0] + [S, H_1] + ...
    \end{aligned}
\end{equation}
where the first $H_1$-term can effectively be discarded by solving
\begin{equation}
    [S, H_0] = -H_1 \ .
\end{equation}
In our case, for a diagonal $H_0 = \Lambda$ and $H_1 = \Tilde{V}$, one finds that the generator $S$ is
\begin{equation}
    S_{ij} = \frac{\Tilde{V}_{ij}}{\lambda_i - \lambda_j}
\end{equation}
where the $\lambda_i$ are the diagonal entries of $\Lambda$. In fact, the transformed Hamiltonian $\Tilde{H}$ up to second order in $\Tilde{V}$ reads
\begin{equation}
    \Tilde{H}_{ij} \approx \Lambda_{ij} + \frac{1}{2} \sum_{k} \Tilde{V}_{ik} \Tilde{V}_{kj} \left( \frac{1}{\lambda_i - \lambda_k} + \frac{1}{\lambda_j - \lambda_k} \right) \ .
\end{equation}
Since $\Tilde{V}_{ij}$ is only different from zero if $i$ and $j$ are in subspaces with distinct $M$, we can see that the leading-order corrections to the Hamiltonian $H_0$ are of order $1/|J|$.\\
The dressed energies and eigenstates are given by
\begin{equation}
    \begin{aligned}
        \Tilde{\lambda}_i &= \lambda_i + \sum_{k\neq i} \frac{\Tilde{V}_{ik} \Tilde{V}_{ki}}{\lambda_i - \lambda_k} \\
        |\Tilde{\psi}_i \rangle &= |\psi_i\rangle + \sum_{k \neq i} \Tilde{\psi}_{ik} |\psi_k \rangle
    \end{aligned}
    \label{eq:SWDressed}
\end{equation}
respectively, where the $|\psi_i\rangle$ are the eigenstates of $\Lambda$ and where we define the correction amplitudes
\begin{equation}
    \Tilde{\psi}_{ik} := \sum_{k'} \Tilde{V}_{kk'} \Tilde{V}_{k'i} \frac{2\lambda_{k'} - \lambda_k - \lambda_i}{(\lambda_{k'} - \lambda_i)(\lambda_{k'} - \lambda_k)(\lambda_i - \lambda_k)}
    \label{eq:Amplitudes}
\end{equation}
for short-hand.\\
As mentioned above, since $\Tilde{V}$ only connects eigenstates in distinct $M$-subspaces, we find $\Tilde{\lambda}_i - \lambda_i = O(|J|^{-1})$. The amplitude $\Tilde{\psi}_{ik} = O( |J|^{-1})$ if $i$ and $k$ are in the same $M$-subspace of the Hamiltonian, while $\Tilde{\psi}_{ik} = O(|J|^{-2})$ if $i$ and $k$ are in different $M$-subspaces.\\
Denote with
\begin{equation}
    \begin{aligned}
        U_T = \exp[-iT\Lambda] = \sum_i e^{-iT\lambda_i} |\psi_i\rangle \langle \psi_i |
    \end{aligned}
\end{equation}
and
\begin{equation}
    \begin{aligned}
        \Tilde{U}_T &= \exp[-iT(\Lambda + \Tilde{V})] = \sum_i e^{-iT\Tilde{\lambda}_i} |\Tilde{\psi}_i \rangle \langle \Tilde{\psi}_i| \\
        &\approx \sum_i e^{-iT\Tilde{\lambda}_i} \Bigg[ \Bigg. |\psi_i\rangle \langle \psi_i | \\
        &+ \sum_{k \neq i} \Tilde{\psi}_{ik} |\psi_k \rangle \langle \psi_i| + \Tilde{\psi}_{ik}^* |\psi_i \rangle \langle \psi_k| \Bigg. \Bigg]
    \end{aligned}
\end{equation}
the time-evolution for a time $T$ under the original Hamiltonian $\Lambda$ and the perturbed Hamiltonian $\Lambda + \Tilde{V}$ up to first-order correction respectively. The correction amplitudes $\Tilde{\psi}_{ik}$ now appear in the off-diagonal elements of $\Tilde{U}_T$ and thus the transition amplitudes between subspaces of different $M$ are suppressed by $O(|J|^{-2})$, i.e. the leakage of probability is $O(|J|^{-4})$.\\
Recall that the eigenstates of $H_{DW}^{(M)}$ and $H_{\text{Fermi}}^{(M)}$ are identical, while their eigenvalues are the same up to a constant energy shift. Therefore, to assess the quality of a simulation of $H_{\text{Fermi}}^{\text{odd}}$ with $H_{DW}$ for a given particle number $M$ it suffices to consider the overlap
\begin{equation}
    \begin{aligned}
        &\langle \Psi_0 | U_T^\dagger \Tilde{U}_T |\Psi_0 \rangle = \sum_i |a_i|^2 e^{iT(\lambda_i - \Tilde{\lambda}_i)} \\ 
        &+ \sum_{k\neq i} a_i^* a_k \left( \Tilde{\psi}_{ik}^* e^{iT(\lambda_i - \Tilde{\lambda}_i)} + \Tilde{\psi}_{ki} e^{iT(\lambda_i - \Tilde{\lambda}_k)}\right)
    \end{aligned}
    \label{eq:SimulationOverlap}
\end{equation}
for some initial state $|\Psi_0 \rangle = \sum_i a_i|\psi_i \rangle$ with support only in the $M$-subspace. The expression in Eq. \eqref{eq:SimulationOverlap} has two contributions, where one is dependent on the correction amplitude $\Tilde{\psi}_{ik}$ and is therefore bounded by $O(1/|J|)$, since we assume $i$ and $k$ to be in the same $M$-block by virtue of initialization. The other contribution, however, is equal to one at $T=0$ and only depends on $J$ via the eigenvalue correction $\lambda_i - \Tilde{\lambda}_i$ and can become small over time. Hence, for $T \leq O(|J|)$ and $|J|$ large, the overlap can be approximated as
\begin{equation}
    \langle \Psi_0 | U_T^\dagger \Tilde{U}_T |\Psi_0 \rangle \approx \sum_i |a_i|^2 e^{i O(T /|J|)} \ .
    \label{eq:ApproxOverlap}
\end{equation}
From this we can conclude that the leading error source when simulating the dynamics of $H_{\text{Fermi}}^{\text{odd}}$ is dephasing at a time scale $O(|J|)$ due to the perturbed eigenenergies of the effective Hamiltonian. The fidelity of the simulation can then be approximated as
\begin{equation}
    \begin{aligned}
        \mathcal{F} &= |\langle \Psi_0 | U_T^\dagger \Tilde{U}_T |\Psi_0 \rangle|^2 \\
        & \approx \sum_i |a_i|^4 \\
        &+ 2 \sum_{j>i} |a_i|^2 |a_j|^2 \cos((\lambda_i - \Tilde{\lambda}_i - \lambda_j + \Tilde{\lambda}_j) T) \\
        & \approx \sum_i |a_i|^4 \\
        &+ 2 \sum_{j>i} |a_i|^2 |a_j|^2 \left( 1 - O(T^2 |J|^{-2}) \right) \\
        &= 1 - O(T^2 |J|^{-2})
    \end{aligned}
    \label{eq:FidelityOverTime}
\end{equation}
where the first approximation is from Eq. \eqref{eq:ApproxOverlap} and the second is using the fact that $\lambda_i - \Tilde{\lambda}_i = O(|J|^{-1})$ and expanding the cosine for small $T/|J|$.\\
We find that the fidelity of the simulation decays quadratically over time $T$. Furthermore, Eq. \eqref{eq:ApproxOverlap} suggests that if the initial state $|\Psi_0\rangle$ is an eigenstate, i.e. all but one $a_i$ are zero, the time evolution is exact up to an irrelevant global phase. If, on the other hand, $|\Psi_0 \rangle$ is highly delocalized in the eigenbasis, the right-hand side of Eq. \eqref{eq:ApproxOverlap} is the weighted sum of many frequencies. This is likely to lead to a faster decay of the simulation fidelity, unless there is some relation between the energy corrections leading to constructive interference. Therefore, we can expect that the time-evolution of initial states close to eigenstates is typically going to be more precise.\\
Note that in the literature it is common to use the spectral norm of two time evolution operators as a metric for the error of a quantum simulation. Here, we consider the infidelity $\varepsilon = 1-\mathcal{F}$. Unlike the spectral norm, the infidelity is insensitive to global phase factors. Since our method inevitably produces global phases due to the energy offsets, the infidelity is the more natural metric to gauge its accuracy. However, the spectral norm could also be considered after correcting for the known global phases and ultimately both errors are equivalent.

\subsection{Parameter re-scaling and time overhead}
The previous analysis shows that the static properties of the original Hamiltonian projected into the $M$-subspace $H_0^{(M)}$ are more closely reproduced by $H_{\text{eff}}^{(M)}$ if the distinct $M$-subspaces are energetically well separated in $\Tilde{H}$. For this to be true, we require $|J|$ to be large in relation to $\| \Tilde{V} \|$. Increasing the coupling strength $|J|$ arbitrarily large in order to obtain a desired fidelity of the simulation might not be an option on many devices. Alternatively, it is possible to leverage precise control of the device. Since $\Tilde{V}$ is given by the off-diagonal elements between subspaces of distinct $M$ induced by the transverse fields in $H_{DW}$ Eq. \eqref{eq:DomainWallHamiltonianFull}, $\| \Tilde{V} \|$ scales linearly with the hopping energies $|t_n|$. We can exploit this to improve the fidelity of the simulation at the cost of moderate time-overhead, as we will discuss now.\\
Let us denote the vector of the parameters $t_n$, $\epsilon_n$ and $v_n$ of $H_{\text{Fermi}}$ collectively as $\mu$ and consider $H_{DW}(\mu, J)$ the domain wall Hamiltonian Eq. \eqref{eq:DomainWallHamiltonianFull} with parameters $\mu$ and domain wall coupling $J$. It is easy to see from Eq. \eqref{eq:DomainWallHamiltonianFull}, that $H_{DW}$ is homogeneous in $\mu$ and $J$, i.e. that for some real $\alpha$
\begin{equation}
    \alpha H_{DW}(\mu, J) = H_{DW}(\alpha \mu, \alpha J) \ .
    \label{eq:HDWHomogeneous}
\end{equation}
Assume a system is evolving according to $H_{DW}(\mu, J)$ for some time $T$, then the time evolution operator will be $\exp[-iTH_{DW}(\mu, J)]$. By dividing the fermi-parameters $\mu$ by some positive number $\alpha$, while simultaneously replacing $T \rightarrow \alpha T$, we find with Eq. \eqref{eq:HDWHomogeneous} that
\begin{equation}
    \alpha T H_{DW}(\mu/\alpha, J) = T H_{DW}(\mu, \alpha J) \ .
    \label{eq:HDWrescaled}
\end{equation}
Consequently, for a time-overhead $\alpha$, $H_{DW}$ effectively simulates the same $H_{\text{Fermi}}$ for time $T$, but with a domain wall coupling multiplied by $\alpha$.\\
For $\alpha \gg 1$, the perturbative expressions Eq. \eqref{eq:SWDressed} and Eq. \eqref{eq:ApproxOverlap} become asymptotically exact, since the generator $S$ of the SW transformation is proportional to $|J|^{-1}$ and, therefore, gets re-scaled by $\alpha^{-1}$ as well. Applying the re-scaling to the expression of the fidelity over time Eq. \eqref{eq:FidelityOverTime}, we find that
\begin{equation}
    \mathcal{F} = 1 - O(T^2 |J|^{-2} \alpha^{-2})
    \label{eq:DynamicFidelity}
\end{equation}
i.e. the re-scaling also reduces the error of the time evolution. Assume the simulation is to be accurate up to an error $\varepsilon$ such that
\begin{equation}
    \mathcal{F} \geq 1 - \varepsilon \ .
\end{equation}
Then $\alpha$ needs to be chosen according to
\begin{equation}
    \alpha = O\left( \frac{T}{|J| \sqrt{\varepsilon}} \right) \ .
    \label{eq:TimeOverhead}
\end{equation}
Since $\alpha$ also re-scales the simulation time, the total time to simulate the evolution according to $H_{\text{Fermi}}$ for time $T$ accurately up to an error $\varepsilon$ is $O(T^2 / \sqrt{\varepsilon})$.\\
Compare this to the error bound of the first-order product formula \cite{Childs_2019}, also known as Trotterization \cite{Heyl_2019}, used to simulate continuous time dynamics on gate-based quantum simulators. The gate complexity scales with $O(T^2/\varepsilon_{\| \cdot \|})$, where $\varepsilon_{\| \cdot \|}$ indicates that the error in Ref.\cite{Childs_2019} is computed in terms of the spectral norm. However, $\varepsilon = O(\varepsilon_{\| \cdot \|}^2)$, as discussed in appendix \ref{sec:AppendixC}. Thus, we find that our approach of simulating continuous time evolution with an analog quantum device displays the same asymptotic behavior with respect to time $T$ and infidelity $\varepsilon$.\\
In the next sections we will numerically verify the discussion of the method. For all numerical results, we use the Python library QuTiP \cite{Qutip1, Qutip2}. Furthermore, for the dynamic simulations, we will employ a standard re-scaling. The reason is that we compare Hamiltonians with different parameters and we need to keep the energy scales of $H_{\text{Fermi}}$ small compared to $|J|$. Therefore, unless stated differently, we will re-scale by
\begin{equation}
    \alpha = \max \{ 1, |t_n|, |\epsilon_n|, |v_n| \} \ .
    \label{eq:StandardScaling}
\end{equation}
While the Hamiltonian is re-scaled $H_{\text{Fermi}} \rightarrow H_{\text{Fermi}} / \alpha$ and the evolution time is re-scaled as $T_{\text{evol}} \rightarrow \alpha T_{\text{evol}}$, we will plot the evolution time $T_{\text{evol}}$ on the time axis, unless stated differently. This means, that the results we present below show the time intervals to be simulated, while already considering the time-overhead required to improve the fidelity. This is of particular interest when considering decoherence of the system.

\section{Static analysis: the SSH-model}
\begin{figure*}
    \centering
    \includegraphics[scale=.55]{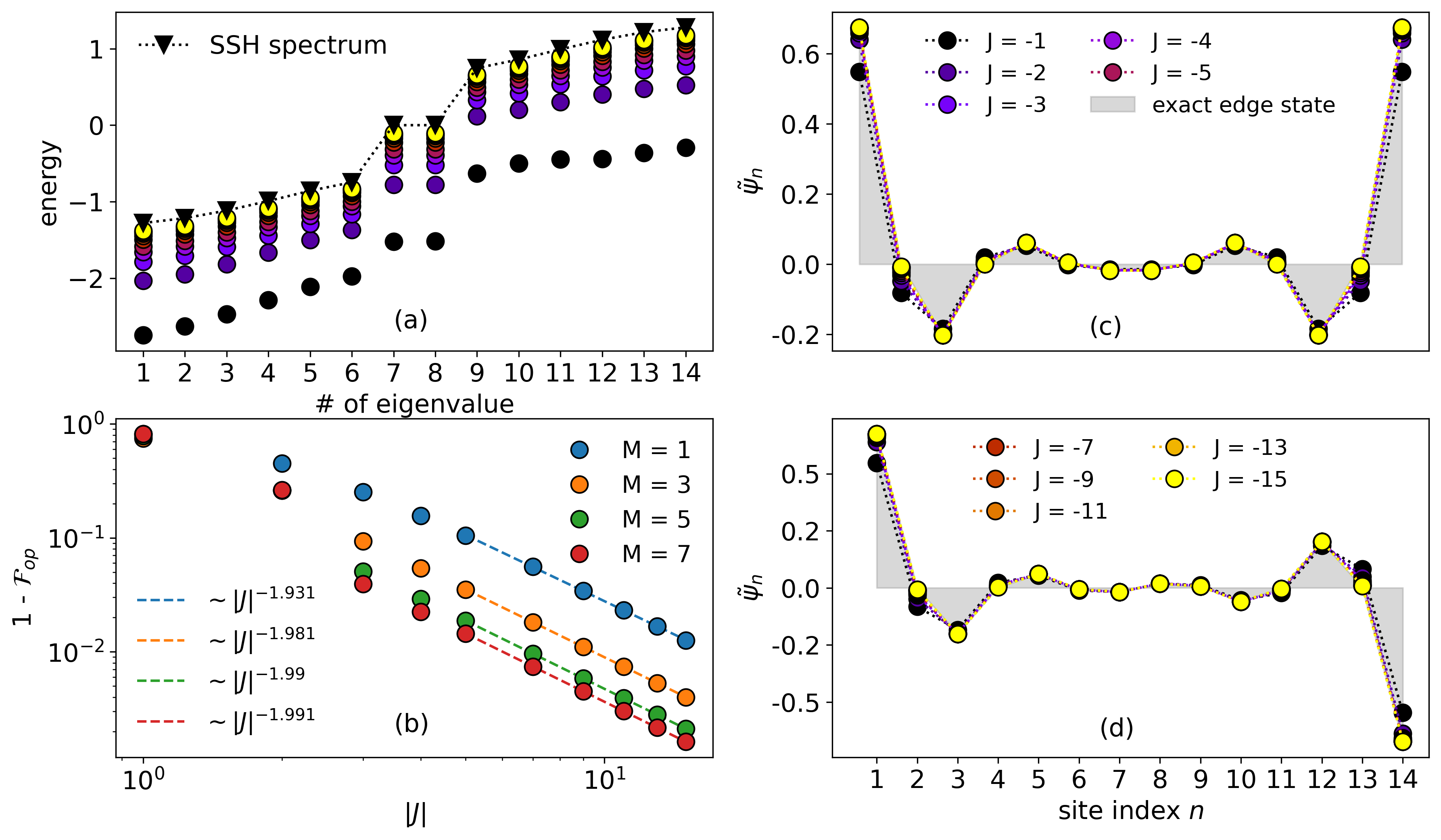}
    \caption{\textbf{(a):} Single-excitation subspace ($M=1$) of $H_{DW}$ simulating $H_{SSH}$ with $v=0.3$, $w=1$ and system size $N=14$ for decreasing $J$; as $|J|$ increases the low-energy spectrum of $H_{DW}$ (circles) approaches the true spectrum of $H_{SSH}$ (triangles) including the eigenvalues 7 and 8 corresponding to the edge states. These states are characteristic of the topologically non-trivial phase; \textbf{(b):} $1-\mathcal{F}_{op}$ over coupling strength $|J|$ in $M$-subspaces ranging from single-particle to half-filling for $N=14$. We fit $\sim |J|^{-\beta}$ to the data points in the asymptotic regime, as depicted. $\mathcal{F}_{op}$ approaches unity for sufficiently large $|J|$. Note that $\beta$ seems to approach 2 as the error decreases with $M$. As discussed in the main text, the improved fidelity as $M$ increases is due to the $M$-subspace being connected to the rest of the Hilbert space by fewer off-diagonal elements of $H_{DW}$ as $M$ approaches $N/2$; \textbf{(c, d):} eigenvector amplitudes of the edge states corresponding to eigenvalue 7 (c) and 8 (d) of $H_{DW}$ for decreasing $J$ in the single-excitation subspace, shaded areas indicating the exact eigenvectors.}
    \label{fig:SSH_Spectrum}
\end{figure*}
We will investigate the static properties of the effective Hamiltonians $H_{DW}^{(M)}$ using the example of the SSH model \cite{Su_1979, Meier_2016}. The SSH model is a one-dimensional model on a linear chain, where the unit cell contains two sites $A$ and $B$. A particle can hop from a site $A$ to site $B$ in either the same or in the previous unit cell. The key feature of the SSH model are the modulated hopping energies within the same unit cell and between two distinct unit cells, denoted by $v$ and $w$ respectively.\\
For simplicity, we identify sites $A$ /$B$ with odd / even sites on a conventional linear chain, which results in the Hamiltonian
\begin{equation}
    H_{SSH} = \sum_{n: \ n \text{\ odd}} v c_{n+1}^\dagger c_n + w c_{n+2}^\dagger c_{n+1} + h.c. \ .
    \label{eq:SSHModel}
\end{equation}
The SSH model is one of the simplest models displaying a topological phase transition. For $|v|>|w|$, the system is in a topological trivial phase with a band gap, while for $|v| < |w|$ the system is topologically non-trivial. In the topologically non-trivial phase,  there are zero-energy edge states located in the band gap. These edge states are localized on the ends of the SSH chain. The two topological phases are separated by a conducting phase $|v| = |w|$. The SSH model has been simulated on quantum simulators based on atom-optics \cite{Meier_2016}, superconducting transmon qubits \cite{Besedin_2021} and quantum dots \cite{Kiczynski_2022}.\\
We numerically prepare the fermionic Hamiltonian $H_{SSH}$ Eq. \eqref{eq:SSHModel} with $w=1$, $v=0.3$ and the corresponding domain wall Hamiltonian $H_{DW}$ Eq. \eqref{eq:DomainWallHamiltonianFull} with the appropriate parameters $\epsilon_n = v_n = 0$ and
\begin{equation}
    t_n = \begin{cases}
        v, & $n$ \text{ odd} \\
        w, & $n$ \text{ even}
    \end{cases} \ .
\end{equation}
In Figure \ref{fig:SSH_Spectrum} (a), we show the spectrum for a single particle. This corresponds to the lowest $N$ eigenvalues of $H_{SSH}$ and $H_{DW}$ respectively. We can clearly identify the bands, as well as the states in the band gap. As the coupling strength $|J|$ increases, the spectrum of $H_{DW}$ approaches the true SSH spectrum.\\
Furthermore, we consider the operator fidelity between the projection of $H_{SSH}$ into the $M$-subspace, denoted by $H_{SSH}^{(M)}$, and the effective Hamiltonian $H_{\text{eff}}^{(M)}$ from Eq. \eqref{eq:DefEffHamiltonian} in Figure \ref{fig:SSH_Spectrum} (b). We define the operator fidelity as the inner product between the normalized $H_{SSH}^{(M)}$ and $H_{\text{eff}}^{(M)}$
\begin{equation}
    \mathcal{F}_{op} = \frac{\langle H_{SSH}^{(M)}, H_{\text{eff}}^{(M)} \rangle^2_F}{ \| H_{SSH}^{(M)} \|_F^2 \| H_{\text{eff}}^{(M)} \|_F^2 }
    \label{eq:DefOperatorFidelity}
\end{equation}
where $\langle \cdot , \cdot \rangle_F$ and $\| \cdot \|_F$ are the Frobenius inner product and norm respectively. As per our perturbative argument above
\begin{equation}
    H_{\text{eff}}^{(M)} = H_{SSH}^{(M)} + O(|J|^{-1})
\end{equation}
and the correction to first order in $|J|^{-1}$ vanishes when expanding $\mathcal{F}_{op}$ around $H_{SSH}^{(M)} = H_{\text{eff}}^{(M)}$. Hence, we expect
\begin{equation}
    \mathcal{F}_{op} = 1 - O(|J|^{-2}) \ .
\end{equation}
In Figure \ref{fig:SSH_Spectrum} (b) we plot $1-\mathcal{F}_{op}$ over $|J|$ for various $M$. We find that in accordance with our perturbative estimates, the fidelity approaches unity at a rate $O(|J|^{-2})$. Interestingly, the fidelity seems to improve for larger values of $M$. At first glance, this is counter intuitive, as the $M$-subspace dimension grows rapidly with $\binom{N}{M}$. However, this observation can be explained by the fact that as $M$ approaches $N/2$, the number of invalid bit flips due to the transverse field driver Hamiltonian is decreasing. For $M$ close to $N/2$, there are more domain walls and a given spin flip is more likely to correspond to a domain wall hopping, rather than an error. In other words, the perturbation operator $V$ in Eq. \eqref{eq:DWHamiltonianDecomposed} is more sparse.\\
The zero-energy states in the SSH spectrum, eigenvalue 7 and 8 in Figure \ref{fig:SSH_Spectrum} (a), correspond to states localized in the edges of the linear chain. We show the amplitudes of the edge states in Figure \ref{fig:SSH_Spectrum} (c) and (d) for increasingly strong domain wall couping $|J|$. The shaded area indicates the amplitudes of the exact eigenstate. We find that even for moderate $|J|$, both eigenvectors are closely approximated. The domain wall is localized in the edges of the chain, which is characteristic of a topological insulating phase.\\
In conclusion, we find that $H_{\text{eff}}^{(M)}$ is an increasingly close approximation of $H_{SSH}^{(M)}$ as $|J|$ grows. In the next section, we will investigate its capability of reproducing dynamical features of non-trivial target Hamiltonians.

\section{Non-interacting system: Aubry-Andre model}
\label{sec:AAModel}
The Aubry-Andre model is defined by the Hamiltonian Eq. \eqref{eq:HoppingHamiltonian}, where the on-site energies are given by
\begin{equation}
    \epsilon_n = \lambda \cos(2\pi \beta n + \phi)
    \label{eq:AAModel}
\end{equation}
while the $v_n = 0$ and, in the simplest case, $t_n = t = 1$. Therefore, this model can be simulated using a simple linear chain of spins and does not require next-nearest-neighbor interactions in $H_{DW}$. For the simulations of a single particle, we will restrict ourselves to homogeneous hopping energies $t_n$. In analogy to the Anderson model, we call $\lambda$ in Eq. \eqref{eq:AAModel} the disorder of the system, but unlike the one-dimensional Anderson model, the Aubry-Andre model exhibits a quantum phase transition (QPT) at $\lambda = 2t$ between a phase of localized eigenstates ($\lambda < 2 t$) and an extended phase ($\lambda > 2 t$), assuming that $\beta$ is irrational \cite{Aubry_1980, Jitomirskaya_1999}. We always set $\beta = \frac{1+\sqrt{5}}{2}$ the golden ratio and $\phi=0$.\\
Besides introducing disorder to the diagonal elements of the Aubry-Andre Hamiltonian, it is also possible to consider disordered off-diagonal terms by setting site-dependent hopping energies $t_n$ according to
\begin{equation}
    t_n = 1+\mu \cos(2\pi \beta (n+1/2) + \phi)
    \label{eq:HoppingDisorder}
\end{equation}
where the parameter $\mu$ quantifies the hopping disorder. This gives rise to a rich phase diagram, which we will explore in the multiple particle setting.\\

\subsection{Single particle}
\begin{figure*}
    \centering
    \includegraphics[scale=.75]{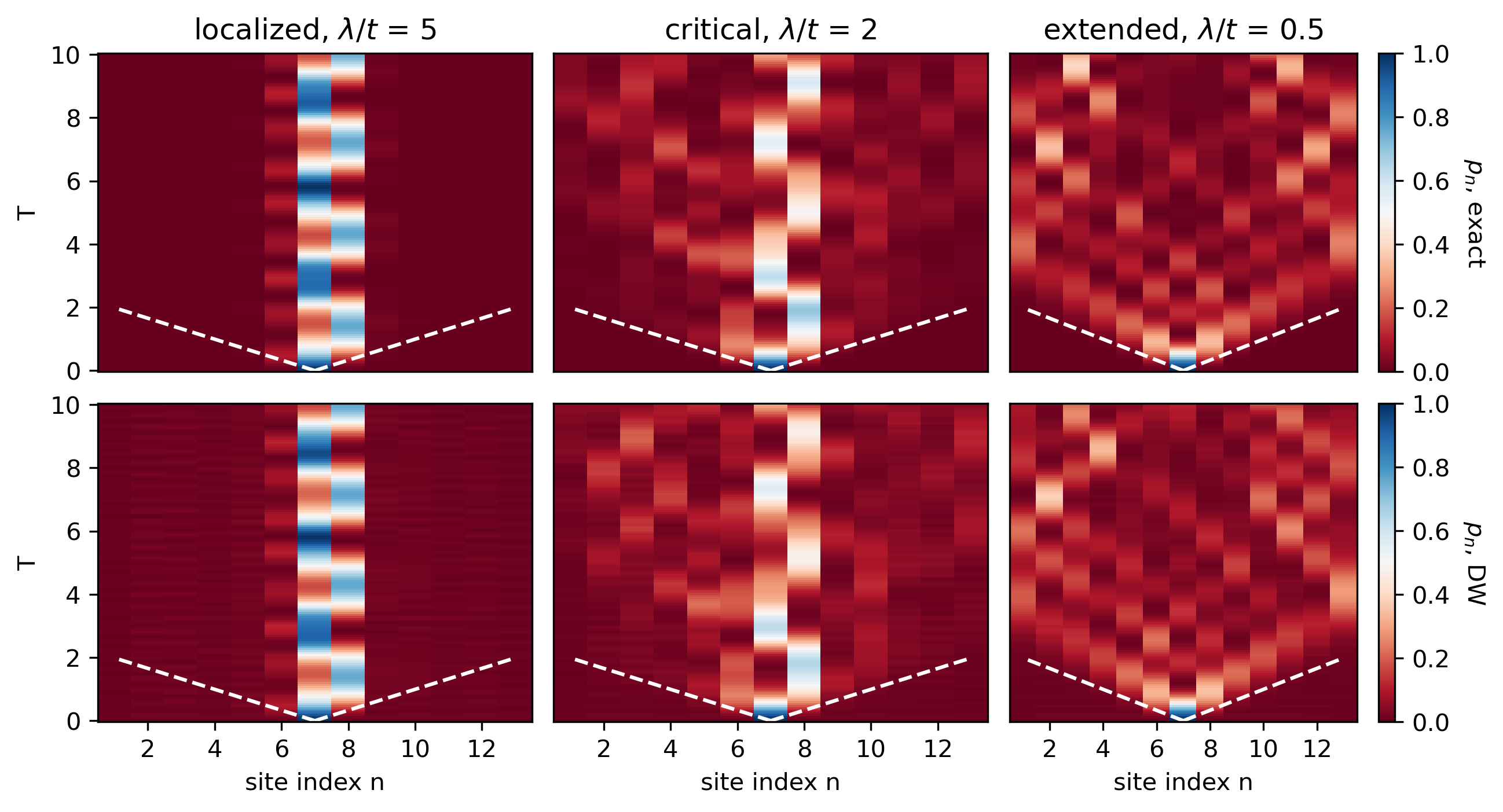}     
    \caption{The simulation results for $t=1$ and domain wall coupling $J=-5$. \textbf{Top row:} Exact time evolution according to the Hamiltonian Eq. \eqref{eq:HoppingHamiltonian} for the Aubry-Andre model with on-site energies Eq. \eqref{eq:AAModel} with $N=13$ sites, \textbf{Bottom row:} time evolution of the domain wall approximation, the state is initialized on site 7, which corresponds to the qubit state $|111111000000\rangle$. The white dashed line shows the Lieb-Robinson bound \cite{Lieb_1972} with $v_{LR} = 3.02$ \cite{Wang_2020}. We do not apply the standard re-scaling Eq. \eqref{eq:StandardScaling}, but instead show the evolution over real time.}
    \label{fig:TimeBoth}
\end{figure*}
\begin{figure*}
    \centering
    \includegraphics[scale=.55]{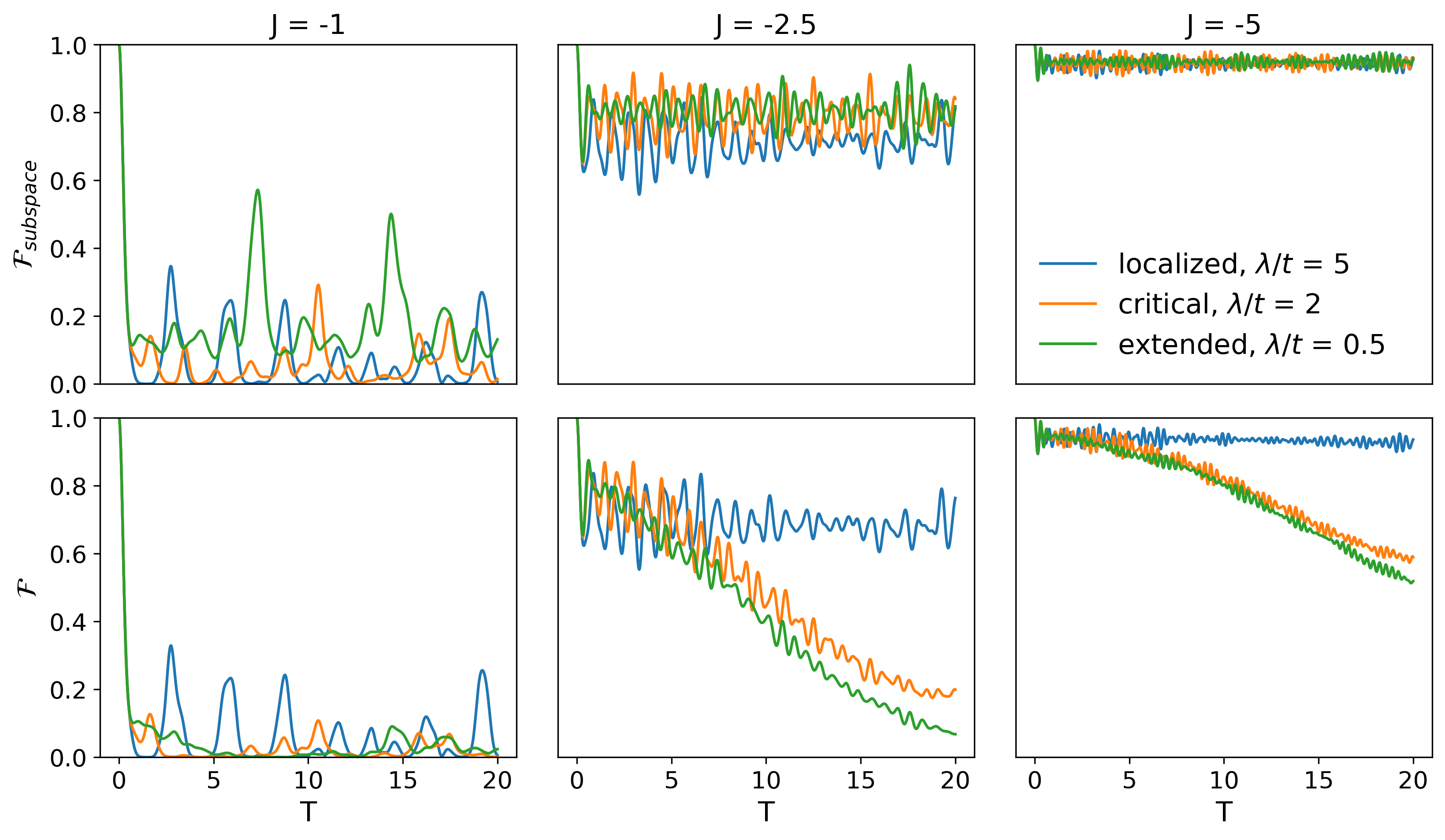}
    \caption{Subspace fidelity $\mathcal{F}_{\text{subspace}}$ \textbf{(top)} and state fidelity $\mathcal{F}$ \textbf{(bottom)} for different values of the coupling $J$. As for the simulation shown in Figure \ref{fig:TimeBoth}, the system is initialized with a single particle on site 7 and evolves under an Aubry-Andre Hamiltonian in the localized, critical and extended phase respectively. As in Figure \ref{fig:TimeBoth}, no re-scaling of the model parameters and time is applied and the results shown are for the real evolution time.}
    \label{fig:FidelitiesSingleParticle}
\end{figure*}

First we compare the time evolution induced by the Hamiltonian $H_{\text{Fermi}}$ and $H_{DW}$ in the respective subspace. We simulate a system of $N=13$ sites, initialize the particle on site 7 and simulate unitary time evolution according to the Hamiltonians $H_{\text{Fermi}}$ and $H_{DW}$. We do not apply the standard re-scaling Eq. \eqref{eq:StandardScaling} in order to obtain an intuition of the impact of the domain wall coupling $|J|$. In Figure \ref{fig:TimeBoth}, we plot the occupations of each site
\begin{equation}
    p_{n,T} = |\langle n | \psi_T \rangle|^2 \ .
\end{equation}
As discussed in more detail in appendix \ref{sec:AppendixB}, in the domain wall model we can measure the $p_n$ at time $T$ via the expectation values
\begin{equation}
    \begin{aligned}
        p_{n, T} = \begin{cases}
            \frac{1}{2} - \frac{1}{2} \langle \sigma_1^z \rangle_T &, n = 1 \\
            \frac{1}{2} + \frac{1}{2} \langle \sigma_{N-1}^z \rangle_T &, n = N\\
            \frac{1}{2} - \frac{1}{2} \langle \sigma_{n-1}^z \sigma_n^z \rangle_T &, \text{else}
        \end{cases}
    \end{aligned}
    \label{eq:SiteOccupationProb}
\end{equation}
where $\langle O \rangle_T$ denotes the expectation value $\langle \psi_T | O | \psi_T \rangle$. The site populations defined in Eq. \eqref{eq:SiteOccupationProb} are also valid in the multi-particle setting. Note that the required expectation values are at most two-local and are thus easily accessible via measurements in the computational basis, e.g. measurement of the persistent current in superconducting flux qubits. By comparing Figure \ref{fig:TimeBoth} top and bottom row, we can see that the time-evolution of the site occupations is accurately captured by the domain wall approximation in both phases of the model, as well as at the critical point.\\
In order to further investigate the fidelity of the simulation and the role of the domain wall coupling $J$, we run the same simulation with different values for $J$. We compute two different figures of merit:
\begin{enumerate}
    \item Subspace fidelity $\mathcal{F}_{\text{subspace}} = ||P_{DW}^{(1)}|\psi_{DW} \rangle||^2$. This measures how much of the population leaks out of the valid domain wall subspace.
    \item State fidelity: the conventional notion of fidelity $\mathcal{F} = |\langle\psi_{\text{exact}} | \psi_{DW} \rangle|^2$.
\end{enumerate}
The listed fidelities are plotted over time in Figure \ref{fig:FidelitiesSingleParticle} for $J=-1, -2.5, -5$. We run the experiment in each phase of the Hamiltonian. For $J=-1$, both fidelities for all quenches decay rapidly, since the energy scale of the original system is larger than $|J|$ and the spin flip term is strong enough so that the system substantially leaks out of the single-particle subspace.\\
Stronger couplings $J=-2.5$ and $J=5$ increase the fidelity measures of the domain wall simulation for longer time-scales. We find that the dynamics remains restricted to the single domain wall subspace, as indicated by $\mathcal{F}_{\text{subspace}}$ close to 1 for all simulations and not visibly decreasing over time. The state fidelity $\mathcal{F}$ for the simulation of the critical and extended system, however, decreases over time. As expected, this decrease is faster for $J=-2.5$.\\
Note that we initialize the system in a completely localized state. Therefore, after a quench to a Hamiltonian that is still localized, the system is initially close to an eigenstate of the Hamiltonian. As we have argued in section \ref{sec:ErrorAnalysis}, this should reduce the error of the simulation. Indeed, the fidelities for $J=-2.5$ and $J=-5$ when quenching to a localized Hamiltonian do not decay below $\approx 0.6$.\\
We can conclude that even at reasonable coupling strengths of $|J| / |t| \approx 5$, the domain wall approximation closely reproduces the dynamics of the Aubry-Andre model for a single particle. The extended simulation in Figure \ref{fig:TimeBoth} bottom row suggest that information can spread several time through the entire system on a time scale with reasonable fidelities and that many interesting properties of the original system's dynamics can still be probed using the domain wall simulator. Next, we investigate the dynamics of multiple domain walls.\\

\subsection{Multiple spinless fermions}
Here we will not only consider different values for the on-site disorder $\lambda$ in Eq. \eqref{eq:AAModel}, but also introduce disorder in the hopping energies $t_n$ via the parameter $\mu$ in Eq. \eqref{eq:HoppingDisorder}. Note that in the following simulations we employ the standard rescaling Eq. \eqref{eq:StandardScaling}.\\
\begin{figure*}
    \centering
    \includegraphics[scale=.55]{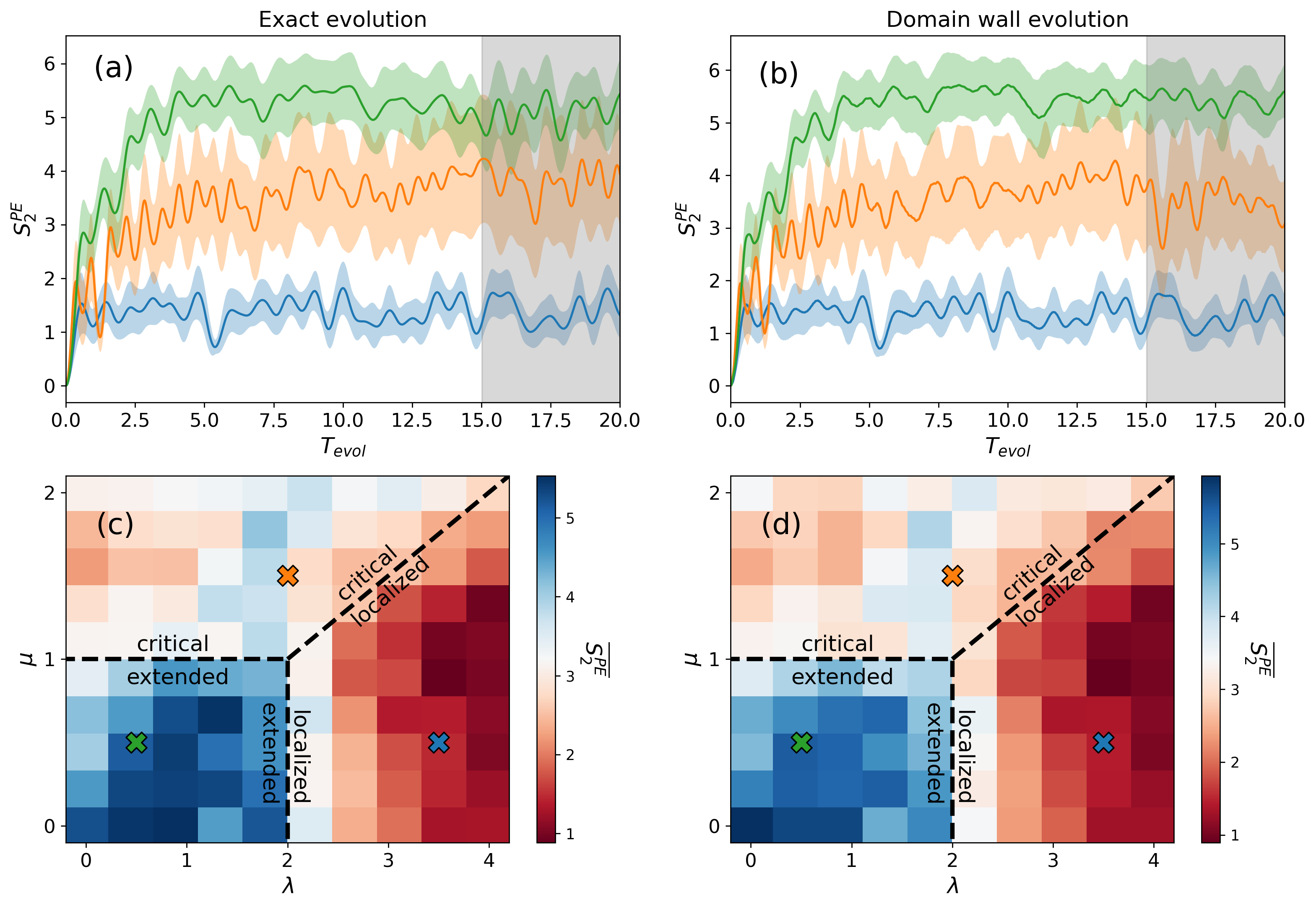}
    \caption{The exact \textbf{(a)} and domain wall approximated \textbf{(b)} participation entropy $S_2^{PE}$ as a function of evolution time $T_{\text{evol}}$ for a system evolving under a localized (blue), critical (orange) and extended (green) Hamiltonian on $N=13$ sites. The solid line and shaded area denote the mean and standard deviation over 10 random initializations with up to $M=\lceil N / 2 \rceil$ particles. The gray shaded interval marks the interval of integration for $\overline{S_2^{PE}}$ used to generate the phase diagrams for exact \textbf{(c)} and approximate \textbf{(d)} evolution. The blue, green and orange crosses correspond to the values of $\lambda$ and $\mu$ for the curves in (a) and (b). The results shown here were obtained using the re-scaling Eq. \eqref{eq:StandardScaling}.}
    \label{fig:AAHalfFillingPD}
\end{figure*}
Following the work by Li et al. \cite{Li_2023}, we can detect the different phases by measuring the second-order participation entropy, i.e. the negative logarithm of the inverse participation ratio (IPR), given by
\begin{equation}
    S_2^{PE} = -\log \sum_{\eta=1}^{\mathcal{N}} |\langle \eta | \psi_\eta \rangle|^4
    \label{eq:DefSPE}
\end{equation}
where the index $\eta$ sums over all $\mathcal{N} = \binom{N}{M}$ computational basis states in the $M$-subspace.\\
We compute $S_2^{PE}$ for each time step as the system evolves over time and average over 10 initializations. Each initialization is a randomly sampled completely localized state of $1 \leq M \leq \lceil N / 2\rceil$ particles. The dynamics of the system is allowed to stabilize for some time and then the time-averaged participation entropy
\begin{equation}
    \overline{S_2^{PE}} =\frac{1}{T_1 - T_0} \int_{T_0}^{T_1} S_2^{PE}(T) dT
\end{equation}
is integrated.\\
Figure \ref{fig:AAHalfFillingPD} (a) and (b) show the time evolution of $S_2^{PE}$ with the solid lines indicating the mean over the 10 initilaizations and the shaded region the standard deviation for the exact evolution and the domain wall approximation respectively. The blue, orange and green curves correspond to the crosses of the same color in the phase diagrams in Figure \ref{fig:AAHalfFillingPD} (c) and (d) respectively. The phase diagrams show $\overline{S_2^{PE}}$ for various values of hopping disorder $\mu$ and onsite disorder $\lambda$. The time interval $[T_0, T_1]$ is indicated as the gray shaded area in Figure \ref{fig:AAHalfFillingPD} (a) and (b).\\
The participation entropies $S_2^{PE}$ obtained from time-evolution of the domain wall model $H_{DW}$ for the localized and extended Hamiltonians closely match the true values almost exactly up to time $\approx 15$. Beyond this time, the approximate $S_2^{PE}$ still seems to match qualitatively. The evolution under the critical Hamiltonian deviates from the true value substantially earlier, but also matches qualitatively.\\
\begin{figure}
    \centering
    \includegraphics[scale=.7]{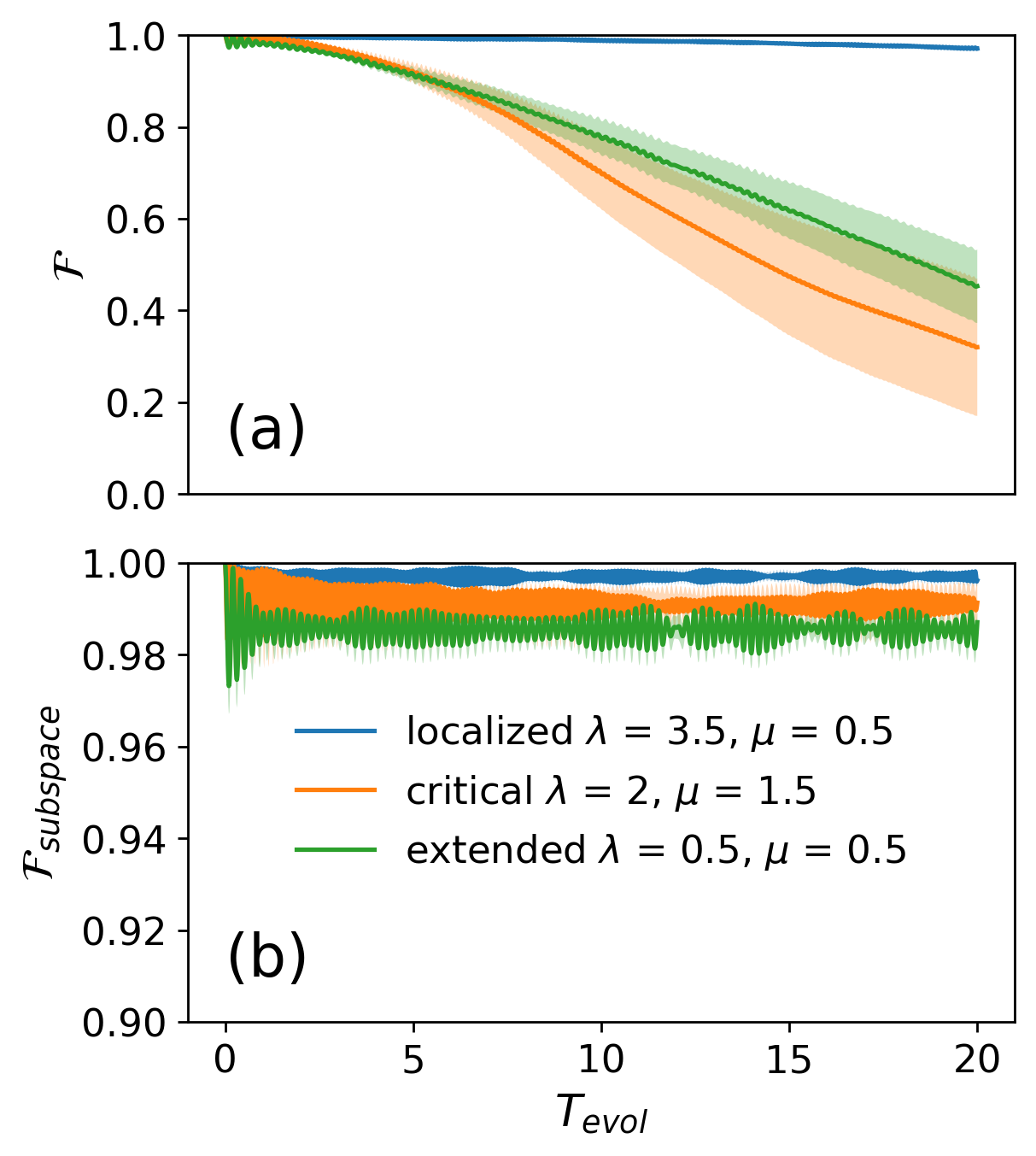}
    \caption{State fidelity $\mathcal{F}$ \textbf{(a)} and subspace fidelity $\mathcal{F}_{\text{subspace}}$ \textbf{(b)} corresponding to the annotated points in the phase diagram in Figure \ref{fig:AAHalfFillingPD} (c) and (d). The solid line is the mean over the 10 random initializations, while the shaded area is the standard deviation.}
    \label{fig:AAHalfFillingFidelities}
\end{figure}
Generally, the approximate time-evolution reproduces the phase diagram of $\overline{S_2^{PE}}$ closely, as can be seen by comparing Figure \ref{fig:AAHalfFillingPD} (c) and (d). This can be further substantiated by investigating the fidelities introduced above. Corresponding to the participation entropies in Figure \ref{fig:AAHalfFillingPD} (a) and (b), the state- and subspace fidelities are shown in Figure \ref{fig:AAHalfFillingFidelities}. As for the single-particle simulations in Figure \ref{fig:FidelitiesSingleParticle}, we find that the evolution is contained predominantly to the correct subspace, while the state fidelity $\mathcal{F}$ decays over the course of the evolution. This decay is the fastest for a system evolving under a critical Hamiltonian.\\
In summary, we find that the domain wall approximation is capable of reproducing the dynamic behaviour of the Aubry-Andre model in all phases. Note that as per our theoretical arguments, as well as the numerics below, the error of the simulations due to the domain wall approximation can be reduced arbitrarily at the cost of some time-overhead. In the next section we will investigate the impact of decoherence on the quality of the simulation.

\subsection{Dynamical quantum phase transition}
Dynamical quantum phase transitions (DQPT) are characterized by non-analytic behaviour in a dynamic observable as a function of time. These non-analytic points, or cusps, have been theoretically analyzed \cite{Zhang_2016} and are considered an analog to the temperature driven phase transitions \cite{Heyl_2013} due to the breakdown of short-time expansions close to said cusps.\\
A common observable used to investigate DQPTs in a system of size $N$ is the rate function $r$ of an initial state $|\psi_0\rangle$ under evolution according to a Hamiltonian $H$ defined by the equation
\begin{equation}
    |\langle \psi_0 | e^{-iH T} | \psi_0 \rangle |^2 = e^{-N r(T)} \ .
    \label{eq:DefRateFunction}
\end{equation}
Usually, DQPTs are observed if the quench dynamics is across a quantum phase transition. We gauge the fidelity by numerically solving the Schrödinger equation both for the the Aubry-Andre model for spinless fermions and domain walls. Additionally, we investigate the impact of decoherence on the quantum simulation by solving the Lindblad master equation
\begin{equation}
    \begin{aligned}
        &\partial_T \rho_T = -i[H, \rho_T] \\
        &+ \sum_i \Gamma_i \left( L_i \rho_T L_i^\dagger - \frac{1}{2} \{ L_i^\dagger L_i, \rho_T \} \right)
    \end{aligned}
\end{equation}
with jump operators $L_i$ and damping rates $\Gamma_i$. The initial condition is $\rho_0 = |\psi_0 \rangle \langle \psi_0|$ and we consider relaxation and dephasing of the individual qubits. The Lindblad operators for a relaxation channel and dephasing are $L_1 = \sigma^-$ and $L_2 = \sigma^z$ respectively applied to each qubit in the simulator. We choose the decay rates $\Gamma_1 = 1/500$ and $\Gamma_2 = 1/100$. Since the jump operators $L_1$ and $L_2$ are breaking the symmetry of the domain wall encoding, to the best of our knowledge, there is no meaningful interpretation of the jump operators in the picture of fermions.\\
\begin{figure*}
    \centering
    \includegraphics[scale=.75]{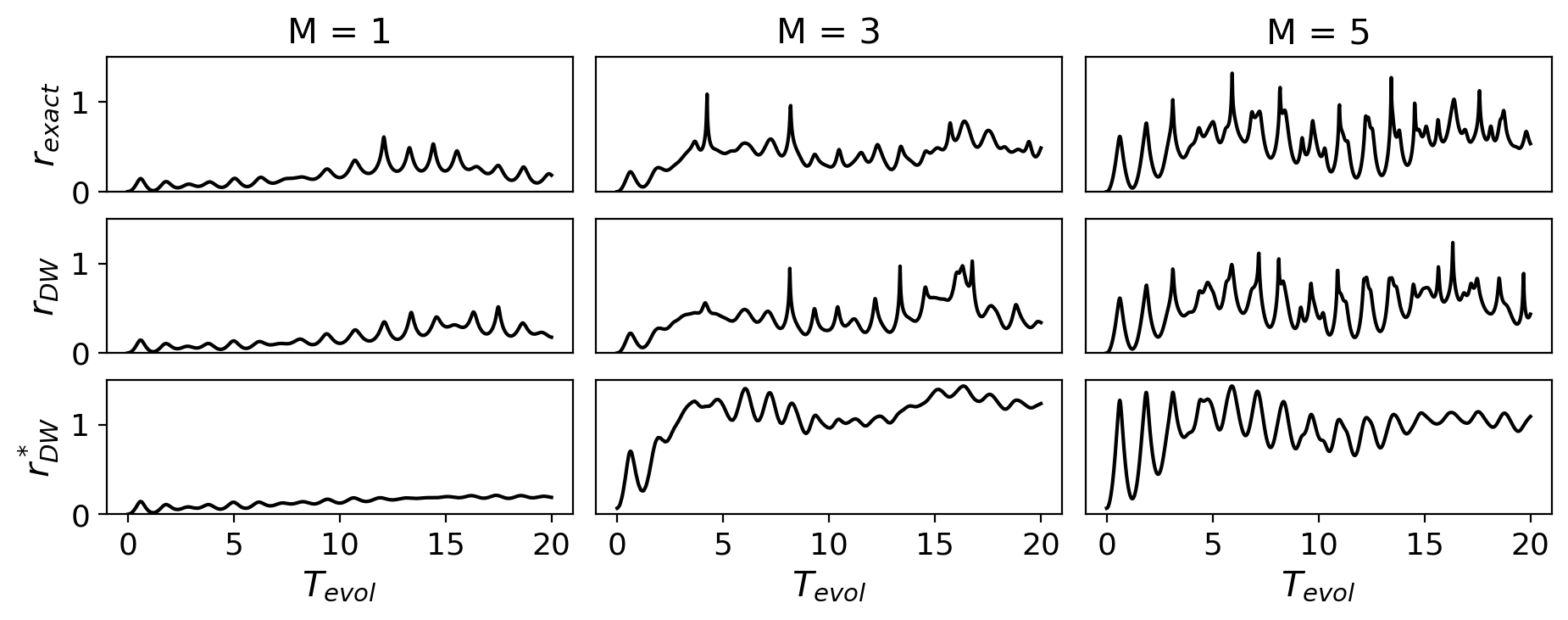}
    \caption{Rate function $r_{\text{exact}}$ for the exact time evolution, the domain wall approximation $r_{DW}$ and the domain wall approximation with noise $r_{DW}^*$; the initial state in Fock space are $|00000100000\rangle$ ($M=1$), $|01000100010\rangle$ ($M=3$) or $|01010101010\rangle$ ($M=5$), while the Hamiltonian is the Aubry-Andre model in the critical phase with $N=11$, $\lambda=2$, $\mu=1.5$; the noise is modeled by relaxation with $\Gamma_1 = 1/500$ and dephasing with $\Gamma_2 = 1/100$ of each individual qubit in the simulator. The results shown here were obtained using the re-scaling Eq. \eqref{eq:StandardScaling}.}
    \label{fig:AAOneParticleNDQPT}
\end{figure*}
\begin{figure*}
    \centering
    \includegraphics[scale=.55]{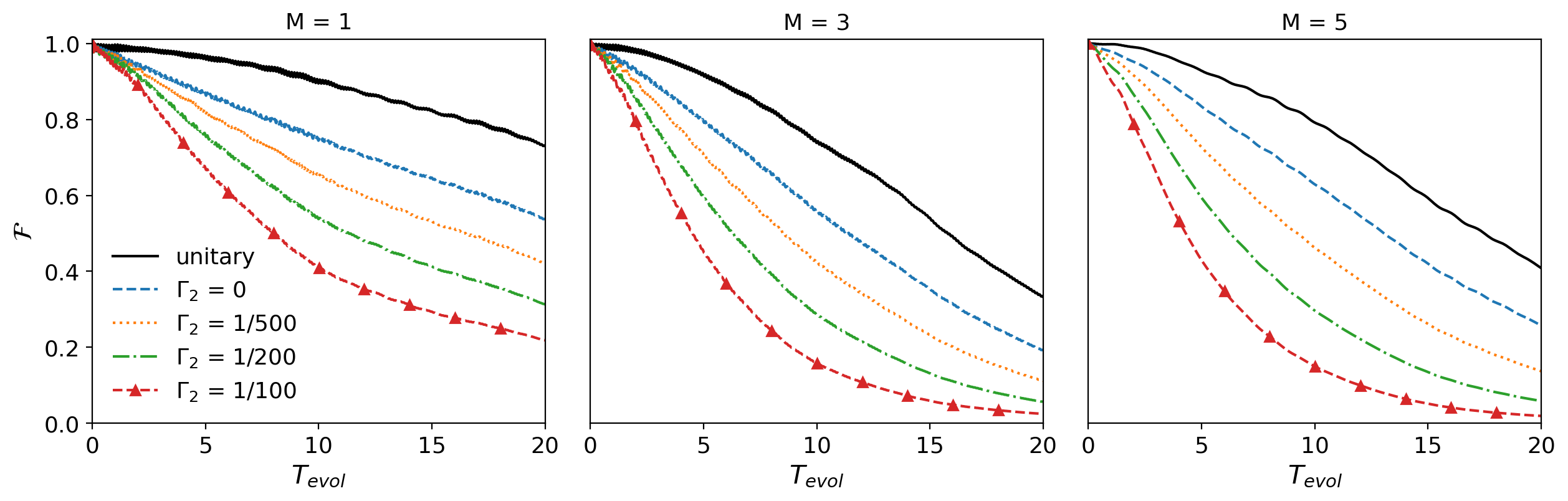}
    \caption{State fidelity $\mathcal{F}$ of the domain wall approximation to the true evolution over time for a unitary evolution of the domain wall approximation, as well as open systems with fixed $\Gamma_1 = 1/500$ and different values of $\Gamma_2$. All other settings are the same as for the rate functions shown in Figure \ref{fig:AAOneParticleNDQPT}. The results shown here were obtained using the re-scaling Eq. \eqref{eq:StandardScaling}.}
    \label{fig:AAOneParticleNDQPT_fidelities}
\end{figure*}
\begin{figure}
    \centering
    \includegraphics[scale=.675]{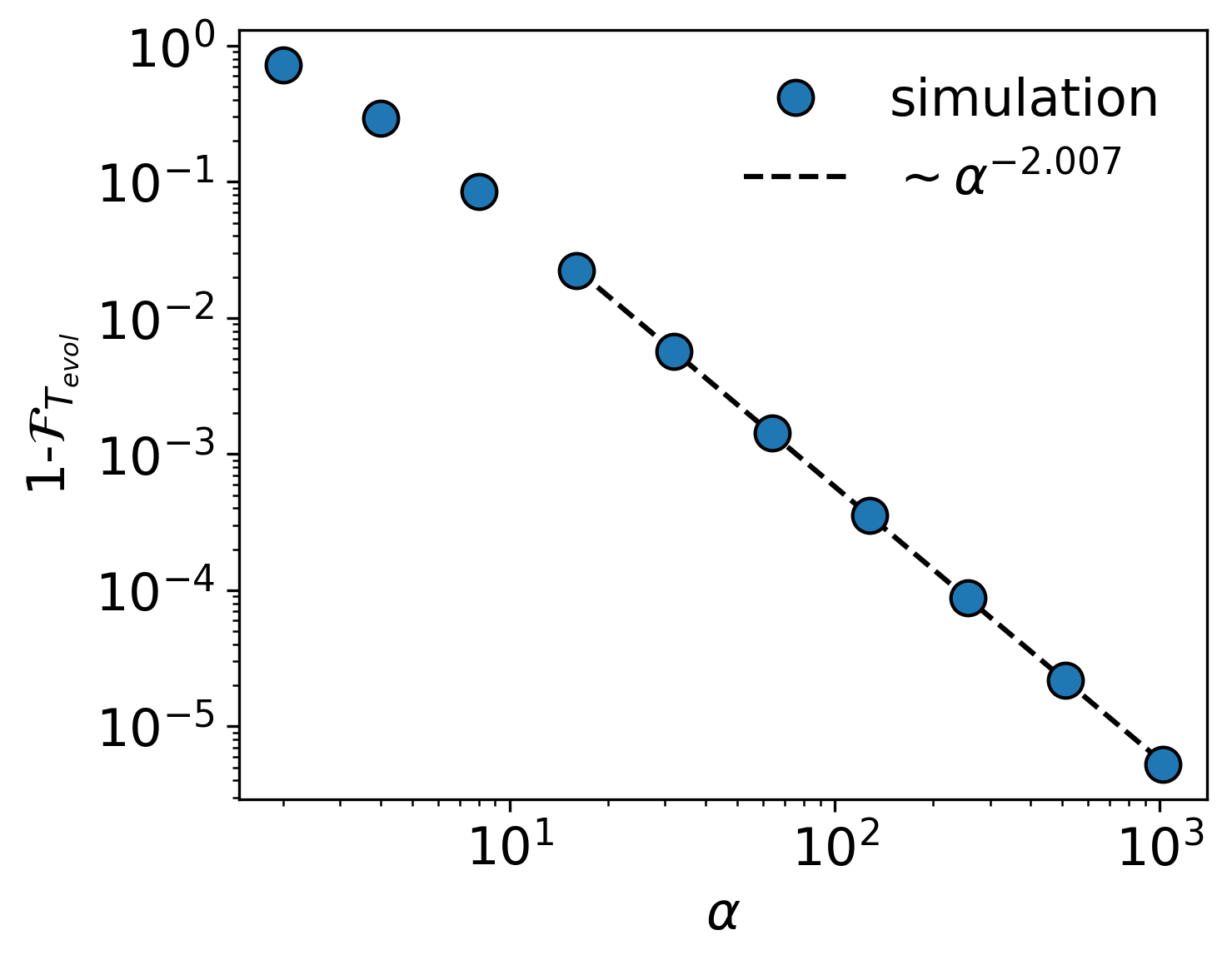}
    \caption{State infidelity $1-\mathcal{F}_{T_{\text{evol}}}$ at the end of the unitary time evolution as a function of the re-scaling factor $\alpha$ Eq. \eqref{eq:HDWrescaled}. The simulation time is $T_{\text{evol}}=20$ for a unitary evolution of the Aubry-Andre model in the same setting as in Figure \ref{fig:AAOneParticleNDQPT} with $M=5$. The wall time of the simulation is $\alpha T_{\text{evol}}$, while the model parameters are divided by $\alpha$. We fit $\alpha^{-\beta}$ to the asymptotic regime, as depicted. In accordance with the analytic argument Eq. \eqref{eq:DynamicFidelity}, the infidelity decays with $\beta \approx 2$.}
    \label{fig:AAErrorScaling}
\end{figure}
The rate functions for the exact evolution $r_{\text{exact}}$, the domain wall approximation $r_{DW}$ and the domain wall approximation in an open quantum system $r_{DW}^*$ are shown in Figure \ref{fig:AAOneParticleNDQPT} for $M=1$, $M=3$ and $M=5$ particles initially uniformly placed over the system. We find that for the single particle, the reproduction of the rate function seems to be of higher quality, but at the same time it shows much less criticality, as seen by the relatively flat rate function. At time $T \approx 12$ there are some cusps in the rate function, but the domain wall evolution seems to capture them only in the closed system, while decoherence seems to smoothen the rate function too much. For $M=3$, the closed system domain wall simulation seems to capture the rate function qualitatively, albeit not exactly. The points in time of the cusps in $r_{DW}$ seem to coincide relatively well with the times of the cusps in $r_{\text{exact}}$, but the sizes differ. In the open system, we find the rate function $r_{DW}^*$ resembles $r_{\text{exact}}$ only for short times $T_{\text{evol}} < 5$. On the other hand, for $M=5$, the behaviour of $r_{\text{exact}}$ is qualitatively well reproduced by $r_{DW}$ and even in the open system, $r_{DW}^*$ seems to capture the correct physics.\\
We can investigate the impact of dephasing in more detail in Figure \ref{fig:AAOneParticleNDQPT_fidelities}, where we show the respective fidelities of the time-evolution under the domain wall Hamiltonian in an open system with the true evolution. Since $\Gamma_1$ is small compared to $\Gamma_2$, we keep it fixed at $\Gamma_1 = 1/500$ and vary the dephasing rate $\Gamma_2$ from no dephasing $\Gamma_2 = 0$ to strong dephasing $\Gamma_2 = 1/100$. Additionally, we show the fidelity of the unitary evolution for comparison. In agreement with the qualitative reproduction of the rate functions in Figure \ref{fig:AAOneParticleNDQPT}, the fidelities for $M=1$ decay the slowest compared to $M=3$ and $M=5$, where the decay seems to be comparable. Apart from a slightly better fidelity for $M=5$ with little to no  dephasing, this holds across all $\Gamma_2$.\\
We find that the simulation using our domain wall approach is rather sensitive to dephasing, as the fidelities in Figure \ref{fig:AAOneParticleNDQPT_fidelities} drop from the unitary evolution quickly as $\Gamma_2$ increases. In the numerical experiments here we have re-scaled the model parameters such that the distinct $M$-subspaces are well separated energetically due to the moderate domain wall coupling $J = -5$. This is sufficient to reproduce the qualitative behaviour of the system, but the unitary evolution is also imperfect and the respective state fidelity decays over evolution time $T_{\text{evol}}$, see Figure \ref{fig:AAOneParticleNDQPT_fidelities}.\\
However, as we have argued theoretically, this fidelity can be improved by increasingly re-scaling both the model parameters, as well as the simulation time. We demonstrate this numerically in Figure \ref{fig:AAErrorScaling}. In case of $M=5$, we compute the unitary time evolution for $T_{\text{evol}} = 20$, as in Figure \ref{fig:AAOneParticleNDQPT} and Figure \ref{fig:AAOneParticleNDQPT_fidelities} and record the infidelity at the end of the evolution for increasing re-scaling parameters $\alpha$, as in Eq. \eqref{eq:HDWrescaled}. As per the theoretical result Eq. \eqref{eq:DynamicFidelity}, the infidelity asymptotically decays with $ \sim \alpha^2$.\\
Since increasing $\alpha$ implies longer simulation time, we expect that there is a trade-off between increased simulation time and therefore improved fidelity of the unitary simulation on one hand, and increased impact of decoherence on the other. The optimal trade-off will depend on the concrete implementation of the simulator.\\
This raises the question how feasible the simulation is on real devices. As stated, this will depend on the concrete implementation, but here we will give some estimates for realistic system parameters. If we identify a unit of energy in our simulation with GHz, then a unit of time corresponds to ns. Hence the experiment discussed in this section requires qubits with relaxation time $T_1 = 500$ ns and dephasing time $T_2 = 100$ ns, which are conservative estimates for contemporary superconducting flux qubits \cite{Yan_2016}. The ferromagnetic coupling energy in the simulations corresponds to 5 GHz, which is also state of the art \cite{HitaPerez_2022}. At the same time, with $\mu = 1.5$ and hopping energies $t_n$ from Eq. \eqref{eq:HoppingDisorder}, the energy splitting of the individual qubits becomes small, if the re-scaling Eq. \eqref{eq:StandardScaling} is applied. Consequently, thermal noise starts to increase with the thermal decay rates \cite{FornDíaz_2017}. While this may not be a severe concern for short simulations, we leave it to further research how this issue can be mitigated. However, we expect that the critical phenomena discussed here can be probed with contemporary quantum hardware.

\section{Interacting system: inhomogeneous XXZ model}
\begin{figure*}
    \centering
    \includegraphics[scale=.5]{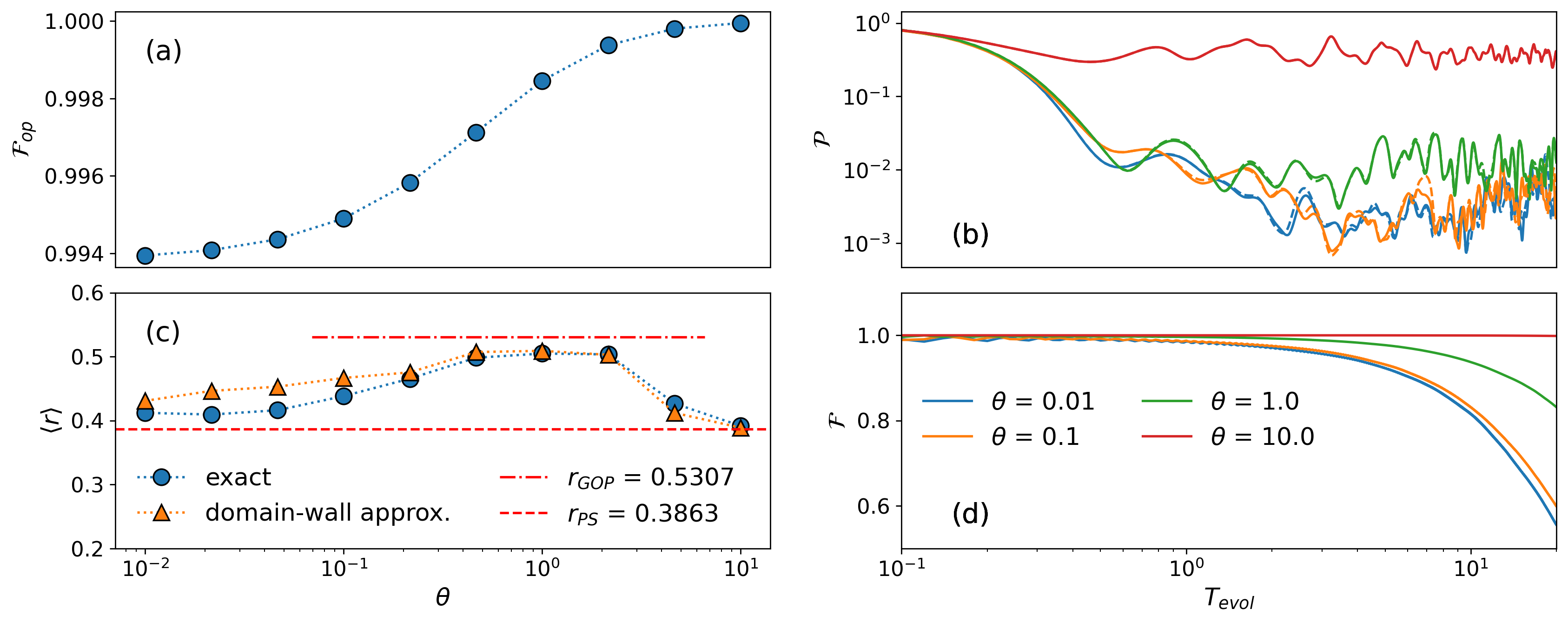}
    \caption{Operator fidelity $\mathcal{F}$ of the true $H_{XXZ}^{(M)}$ and the effective domain wall Hamiltonian $H_{\text{eff}}^{(M)}$ \textbf{(a)} for $N=13$ sites with $\Delta = 1$ and projected onto the $M=7$ particle subspace, as well as the average consecutive level spacing ratio shown in \textbf{(c)} as a function of the inhomogeneity $\theta$. The red dashed and dotted-dashed lines show $\langle r \rangle$ for the Poisson and Wigner-Dyson distributions respectively. Survival probability $\mathcal{P}$ \textbf{(b)} over time for the domain wall approximation (solid lines) and the exact evolution (dashed lines) averaged over 10 random initializations of the system in a product state polarized in Z-direction and the corresponding fidelities \textbf{(d)} for various inhomogeneities $\theta$. The results shown here were obtained using the standard re-scaling Eq. \eqref{eq:StandardScaling}.}
    \label{fig:XXZ1}
\end{figure*}

\begin{figure*}
    \centering
    \includegraphics[scale=.55]{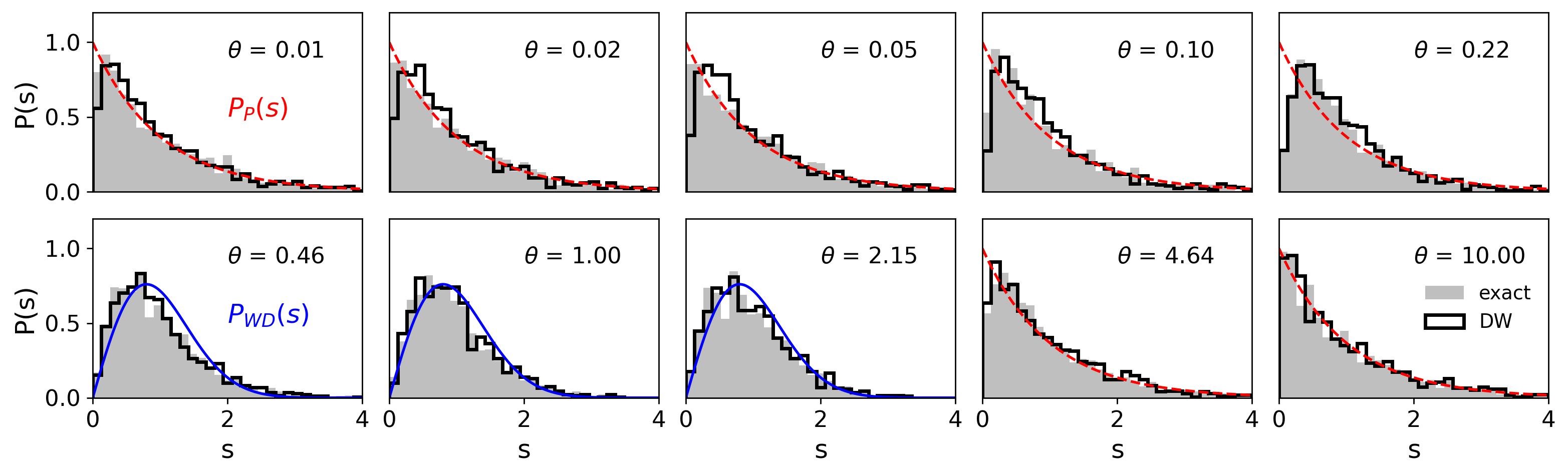}
    \caption{Histograms of the unfolded level spacing $s$ for the exact XXZ model $H_{XXZ}^{(M)}$ (black line) and the effective domain wall Hamiltonian $H_{\text{eff}}^{(M)}$ (shaded gray) with $N=13$ sites after projecting into the $M=7$ excitation subspace for various values of the inhomogeneity $\theta$. Depending on the log-likelihood of the true histogram with respect to the distributions $P_P(s)$ and $P_{WD}(s)$, we show $P_P(s)$ or $P_{WD}(s)$ respectively. As before, the results shown here were obtained using the re-scaling Eq. \eqref{eq:StandardScaling}.}
    \label{fig:XXZ2}
\end{figure*}

The models we have analyzed so far only required hopping and on-site energies, but no interactions between neighboring particles. In order to validate the implementation of the particle interactions, we simulate the inhomogeneous Heisenberg XXZ model.\\
Recently, it has been shown that the degree of the inhomogeneity of the ZZ-interactions impacts the thermalization behaviour of the model \cite{Wang_2024}. Following reference \cite{Wang_2024}, we investigate the XXZ model
\begin{equation}
    \begin{aligned}
        H_{XXZ} &=  \sum_{n=1}^{N-1} t (\sigma_n^x \sigma_{n+1}^x + \sigma_n^y \sigma_{n+1}^y ) + \Delta_n \sigma_n^z \sigma_{n+1}^z \\
        &= \sum_{n=1}^{N-1} 2t (\sigma_n^+ \sigma_{n+1}^- + \sigma_n^- \sigma_{n+1}^+ ) + \Delta_n \sigma_n^z \sigma_{n+1}^z
    \end{aligned}
\end{equation}
where
\begin{equation}
    \Delta_n = \Delta + \theta \frac{2n-N}{N-2} \ .
\end{equation}
The hopping energy is fixed to $t=1$ and $\theta$ is a measure for the inhomogeneity. This model falls into the class of spin chains Eq. \eqref{eq:HSpinChain} that can be simulated with our domain wall approach.\\
Depending on $\Delta$ and $\theta$, the XXZ model can be in an integrable phase, or in a chaotic phase. These two phases can be distinguished by considering the probability distribution of the energy level spacings $P(s)$, where $P(s)$ is a Poisson distribution
\begin{equation}
    P_P(s) = e^{-s}
\end{equation}
in the integrable phase, while in the chaotic phase $P(s)$ is a Gaussian orthogonal ensemble (GOE) following the Wigner-Dyson distribution
\begin{equation}
    P_{WD}(s) = \frac{\pi s}{2} e^{-\frac{\pi s^2}{4}} \ .
\end{equation}
The different distributions result in distinct average ratios of consecutive level spacing
\begin{equation}
    \langle r \rangle = \frac{1}{\mathcal{N}} \sum_{\eta=2}^{\mathcal{N}-1} \min \{ r_{\eta}, 1/r_{\eta} \}
\end{equation}
where
\begin{equation}
    r_{\eta} = \frac{E_{\eta + 1} - E_{\eta}}{E_{\eta} - E_{\eta - 1}} \ .
\end{equation}
We will investigate how well the domain wall approximation captures the level statistics by comparing $\langle r \rangle$ for the exact projection the $M$-subspace $H_{XXZ}^{(M)}$ and the corresponding $H_{\text{eff}}^{(M)}$. Additionally, we compute the operator fidelity $\mathcal{F}_{op}$ for $H_{XXZ}^{(M)}$ and $H_{\text{eff}}^{(M)}$. The results are shown in Figure \ref{fig:XXZ1} (a) and (c).\\
In Figure \ref{fig:XXZ1} (a) we find that $\mathcal{F}_{op}$ is high for all values of inhomogeneity $\theta$, however, there is a slight drop for small $\theta$. Corresponding to the drop in $\mathcal{F}_{op}$, the level spacing $\langle r \rangle$ of the domain wall approximation deviates from the exact value in Figure \ref{fig:XXZ1} (c). This deviation can be understood by considering the level spacing distribution of the unfolded energy levels \cite{Santos_2010, Sierant_2019, Wang_2024} in Figure \ref{fig:XXZ2}. Together with the histograms of the level statistics of $H_{XXZ}^{(M)}$ and $H_{\text{eff}}^{(M)}$, we plot the density $P_P(s)$ or $P_{WD}(s)$, depending on which distribution has the largest log-likelihood relative to the histogram of the exact model.\\
The energy eigenvalues are unfolded as a means to normalize the mean level spacing to one. The unfolding procedure closely follows reference \cite{Wang_2024}: we fit a 12th-order polynomial $\Tilde{N}(E)$ to the cumulative spectral function
\begin{equation}
    N(E) = \sum_n \Theta(E-E_n)
\end{equation}
where $\Theta$ is the Heaviside step function. We compute the normalized eigenvalues
\begin{equation}
    \Tilde{E}_n = \Tilde{N}(E_N)
\end{equation}
and compute the level spacings
\begin{equation}
    s_n = \Tilde{E}_{n+1} - \Tilde{E}_n \ .
\end{equation}
The $s_n$ are binned in a histogram with bin width 1/8. For robustness, we disregard the 200 smallest and largest eigenvalues of the Hamiltonian in this procedure.\\
The exact statistics in Figure \ref{fig:XXZ2} follow the Poisson distribution for small and large $\theta$, while for intermediate $\theta$ the $P_{WD}(s)$ best describes the data, as was discussed in \cite{Wang_2024}. The statistics of the domain wall approximation match this observation, except for small $\theta$. Here, the histogram of the approximation is biased towards larger level spacings, which matches the deviation of $\langle r \rangle$ in Figure \ref{fig:XXZ1} (c). This is best explained as a consequence of the $O(|J|^{-1})$-perturbation in $H_{DW}^{(M)}$, as discussed in section \ref{sec:ErrorAnalysis}. The perturbation introduces additional off-diagonal elements, which cause level repulsion and thus broaden the level spacing statistics.\\
In order to measure the quality of the reproduction of the dynamic properties of the model, we compute the survival probability of an initial states $|\psi_0\rangle$ when the system evolves according to $H = H_{DW}, H_{XXZ}$
\begin{equation}
    \mathcal{P} = |\langle \psi_0 | e^{ -i H T} | \psi_0 \rangle |^2
\end{equation}
as well as the fidelity of the simulation
\begin{equation}
    \mathcal{F} = |\langle \psi_0 | e^{ i H_{XXZ} T} e^{ -i H_{DW} T} | \psi_0 \rangle |^2 \ .
\end{equation}
In line with the high operator fidelities in Figure \ref{fig:XXZ1} (a), the dynamical properties are well reproduced. The survival probability $\mathcal{P}$ of the domain wall approximation coincides closely with the solution of the XXZ model, as seen in Figure \ref{fig:XXZ1} (b), while the fidelities in Figure \ref{fig:XXZ1} (d) remain $>0.5$ throughout the simulation. The fidelity of the simulation increases with $\theta$. This is a consequence of the standard energy re-scaling Eq. \eqref{eq:StandardScaling} to effectively increase the domain wall coupling, such that the $M$-subspaces of $H_{DW}$ remain well separated even for large $\theta$. Note, that by using the standard re-scaling Eq. \eqref{eq:StandardScaling} the fidelities can be improved arbitrarily at a linear time-overhead.

\section{Time-dependent Hamiltonian: Next-nearest-neighbor hopping via Floquet engineering}
We have shown that a simple Ising-type Hamiltonian with strong coupling can be used to simulate spinless fermions in one dimension. The Hamiltonians we have simulated so far were all only using nearest-neighbor hopping terms. If the model parameters can be controlled and varied over time, it is possible to overcome this limitation to some degree using Floquet engineering. In this section, we will argue that using time-dependent hopping energies $t_n$ it is possible to simulate imaginary next-nearest-neighbor hopping \cite{Novo_2021}.\\
Closely following reference \cite{Novo_2021}, we review the basic Floquet theory and its application to the problem at hand. Starting out with a simple Hamiltonian of spinless fermions in one dimension
\begin{equation}
    H_{NN} = - \sum_n t_n c_{n+1}^\dagger c_n + h.c.
    \label{eq:HFloquet}
\end{equation}
we can vary the hopping energies as functions of time $T$ with period $\tau$
\begin{equation}
    t_n(T) = t_n(T+\tau)
\end{equation}
and thus make the Hamiltonian $H_{NN}(T) = H_{NN}(T+\tau)$ also periodic. The key realization of Floquet engineering is that the time evolution according to $H_{NN}$ over one period is given by a Floquet Hamiltonian $H_{F}$
\begin{equation}
    \mathcal{T} \exp \left[ -i\int_0^\tau H_{NN}(T) dT \right] = \exp \left[ -i \tau H_{F} \right] \ .
\end{equation}
The time evolution for an integer multiple of the period $T_{\text{evol}} = m\tau$ is described by
\begin{equation}
    \mathcal{T} \exp \left[ -i\int_0^{T_{\text{evol}}} H_{NN}(T) dT \right] = \exp \left[ -i \tau H_F \right]^m \ .
\end{equation}
$H_F$ is a time-independent Hamiltonian that can have very different properties from the original $H_{NN}$. Generally, $H_F$ can only be computed approximately by using the truncated Magnus expansion
\begin{equation}
    H_F = \sum_{p=0}^{\infty} H_F^{(p)} \approx H_F^{(0)} + H_F^{(1)}
\end{equation}
where the first two terms are given by
\begin{equation}
    H_F^{(0)} = \frac{1}{\tau} \int_0^{\tau} H(T) dT
\end{equation}
\begin{equation}
    H_F^{(1)} = \frac{1}{2i\tau} \int_0^{\tau} \int_0^{T} \left[ H(T), H(T') \right] dT' dT
\end{equation}
respectively. Considering the Fourier decomposition of $H(T)$ in multiples of the frequency $\Omega = 1/\tau$
\begin{equation}
    H(T) = \sum_{l=-\infty}^\infty H_l e^{i l \Omega T}
\end{equation}
the terms of the Magnus expansion read
\begin{equation}
    H_F^{(0)} = H_0
\end{equation}
\begin{equation}
    H_F^{(1)} = \frac{1}{\Omega}\sum_{l=1}^\infty \frac{1}{l} (\left[ H_l, H_{-l} \right] - \left[ H_l, H_0 \right] + \left[ H_{-l}, H_0 \right]) \ .
\end{equation}
Choosing the $t_n(T)$ as superpositions of two cosines
\begin{equation}
    \begin{aligned}
    t_n(T) = &t_n^{(0)} + t_n^{(1)} \cos(\Omega T + \phi_n^{(1)}) \\
    &+ t_n^{(2)} \cos(2\Omega T + \phi_n^{(2)})
    \end{aligned}
\end{equation}
the terms of the Fourier series are found to be
\begin{equation}
    H_l = \sum_n t_n^{(l)} e^{ i \phi_n^{(l)} l / |l|} (c_n^\dagger c_{n+1} + c_{n+1}^\dagger c_n)
\end{equation}
for $l= \pm 1, \pm 2$ and $H_l = 0$ otherwise. Choosing the amplitudes and phases of the drive signal as
\begin{equation}
    \begin{aligned}
        &t_n^{(0)} = -\frac{3K_1\tau}{4\pi} \ \ ,\\
        &t_n^{(1)} = \sqrt{|K_2|} \ \ , \ \ \phi_n^{(1)} = -n\frac{\pi}{2}\\
        &t_n^{(2)} = 2t_n^{(1)} \ \ , \ \ \phi_n^{(2)} = \phi_n^{(1)} + \pi \\
    \end{aligned}
\end{equation}
one finds after some algebraic manipulation that the Floquet Hamiltonian $H_F$ is proportional to a Hamiltonian with nearest-neighbor and imaginary next-nearest-neighbor hopping
\begin{equation}
    \begin{aligned}
    \frac{4\pi}{3\tau}H_F &= H_{NNN} \\
    &= \sum_n K_1 c_{n+1}^\dagger c_n + iK_2 c_{n+2}^\dagger c_n + h.c. \ .
    \end{aligned}
    \label{eq:Heff}
\end{equation}
Due to the proportionality factor, the simulation time has to be chosen as
\begin{equation}
    T_{\text{sim}} = \frac{4\pi T_{\text{evol}}}{3\tau}
    \label{eq:DefFloqTsim}
\end{equation}
so that
\begin{equation}
    T_{\text{sim}} H_F = T_{\text{evol}} H_{NNN} \ .
\end{equation}
Due to the truncation of the Magnus expansion, the Floquet Hamiltonian only describes the time evolution approximately and for a evolution time $T_{\text{evol}}$ the error in terms of the infidelity is
\begin{equation}
    \varepsilon_{\text{Floquet}} \leq O(\tau^4 T_{\text{evol}}^2) \ .
    \label{eq:FLOError}
\end{equation}
For the expression of the infidelity in Eq. \eqref{eq:FLOError} we use the error in terms of the spectral norm from Ref. \cite{Novo_2021}, Eq. (22), and adapt it in accordance with the discussion in appendix \ref{sec:AppendixC}.\\
Clearly, the approximation is more accurate if the period $\tau$ is small. This error is on top of the error due to the domain wall approximation from Eq. \eqref{eq:FidelityOverTime}. We shall denote this error as $\varepsilon_{DW}$ in order to distinguish it from $\varepsilon_{\text{Floquet}}$. Recall that we can control $\varepsilon_{DW}$ by re-scaling the model parameters $t_n$ while keeping $J$ constant and hence
\begin{equation}
    \varepsilon_{DW} = O(T_{\text{sim}}^2 |\alpha J|^{-2}) \ .
    \label{eq:DWError}
\end{equation}
The re-scaling also applies for time-dependent parameters. However, if we maintain the original frequency $\Omega$, we effectively simulate the same time evolution, but with a larger frequency $\alpha \Omega$ and the truncation error of the Magnus expansion is reduced. However, at the same time, we also scale the factor in the first equality of Eq. \eqref{eq:Heff}. Combining these contributions, the infidelity of the simulation for $\alpha > 1$ is
\begin{equation}
    \varepsilon_{\text{Floquet}} = O(\tau^4 \alpha^{-2} T_{\text{evol}}^2) \ .
    \label{eq:FLOError2}
\end{equation}
In order to simulate $H_{NNN}$ up to a precision $\varepsilon$, we demand
\begin{equation}
    \varepsilon_{\text{Floquet}} + \varepsilon_{DW} \leq \varepsilon
\end{equation}
which can be achieved by
\begin{equation}
    \varepsilon_{\text{Floquet}}, \varepsilon_{DW} \leq \varepsilon / 2 \ .
    \label{eq:ErrorCondition}
\end{equation}
Thus, by considering Eq. \eqref{eq:DefFloqTsim}, Eq. \eqref{eq:DWError} and Eq. \eqref{eq:FLOError2}, we can make the simulation accurate up to an error $\varepsilon$ by choosing
\begin{equation}
    \begin{aligned}
        \alpha = O \left( \max \left\{ \frac{ \tau^2 T_{\text{evol}}}{\sqrt{\varepsilon}} , \frac{T_{\text{evol}}}{ \tau |J| \sqrt{\varepsilon}} \right\} \right) \ .
    \end{aligned}
    \label{eq:ScalingFloquet}
\end{equation}
From Eq. \eqref{eq:ScalingFloquet} it is clear that the dominant error depends on the frequency and the domain wall coupling $|J|$. Note that this re-scaling is useful for quantum devices where the drive frequency $\Omega = \tau^{-1}$, the coupling $|J|$ or both cannot be increased arbitrarily.
In summary, the evolution under $H_{NNN}$ for time $T_{\text{evol}}$ can be simulated to arbitrary precision $\varepsilon$ using the Ising-type Hamiltonian $H_{DW}$ with modulated transverse fields in time at most $O(T_{\text{evol}}^2 / \sqrt{\varepsilon})$.\\
\begin{figure*}
    \centering
    \includegraphics[scale=.6]{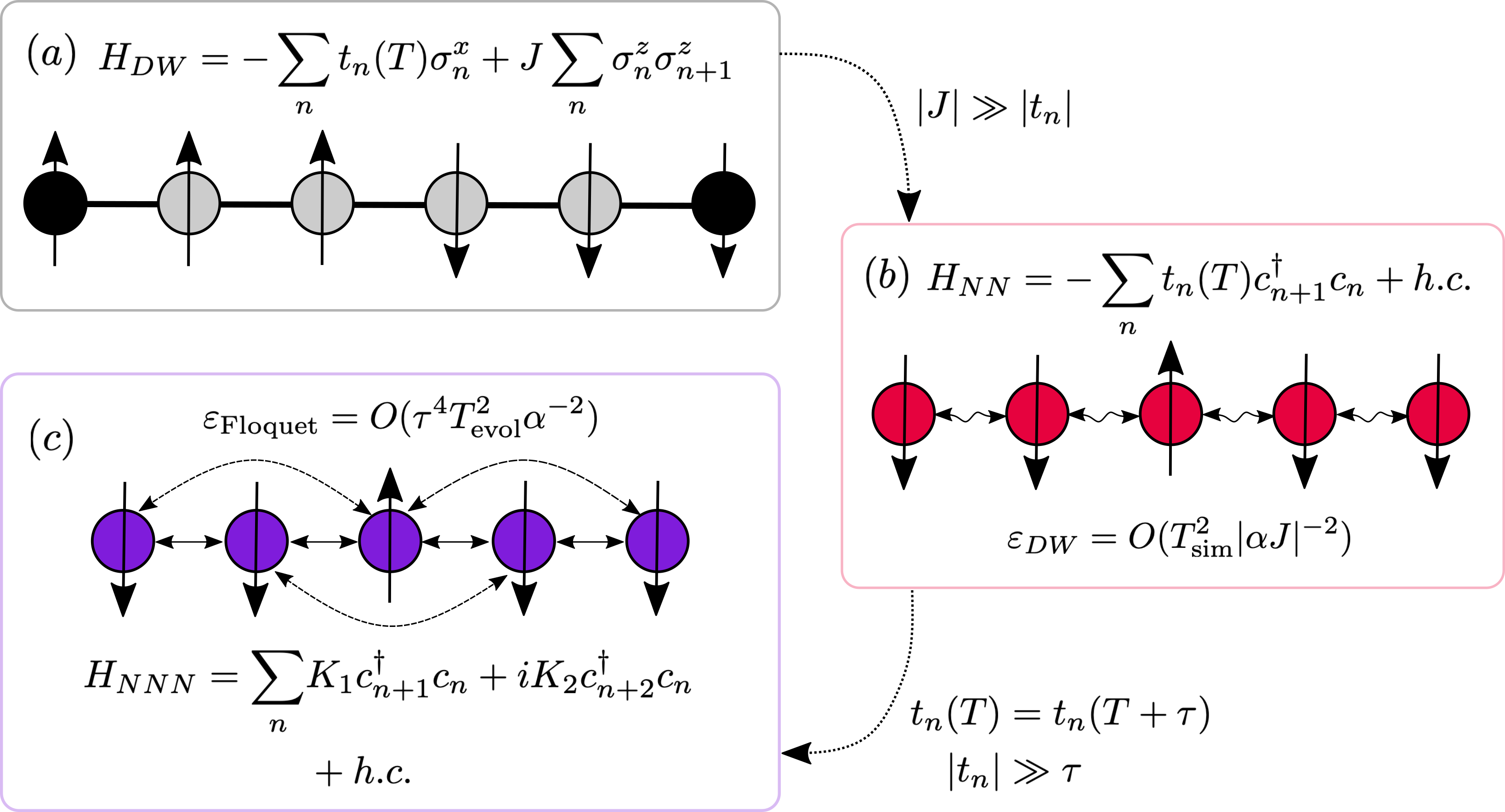}
    \caption{Illustration of the relationships of the distinct models: the Ising Hamiltonian $H_{DW}$ \textbf{(a)} simulates nearest-neighbor hopping $H_{NN}$ \textbf{(b)} if the domain wall coupling is strong enough. Via a periodically driven hopping amplitudes $t_n(T) = t_n(T+\tau)$ with period $\tau$, an effective Hamiltonian $H_{NNN}$ \textbf{(c)} with imaginary next-nearest-neighbor hopping can be simulated. Each approximation layer introduces an error, which can be controlled either by increasing the domain wall coupling strength $|J|$ and the drive frequency $\Omega = 1/\tau$, or by re-scaling the Hamiltonian parameters and incuring a time-overhead $\alpha$, resulting in a total runtime of $O(T_{\text{evol}}^2 / \sqrt{\varepsilon})$.}
    \label{fig:FloquetSchema}
\end{figure*}
Figure \ref{fig:FloquetSchema} depicts the relationship between the different layers of simulation: the Ising-type Hamiltonian $H_{DW}$ with strong coupling $|J|$ and periodically driven transverse fields $t_n(T) = t_n(T+\tau)$ simulates a fermionic Hamiltonian $H_{NN}$ with nearest-neighbor hopping and modulated hopping energies. If the driving is engineered appropriately, $H_{NN}$ effectively realizes a time-independent fermionic Hamiltonian with imaginary next-nearest-neighbor hopping.\\
We compare the time evolution of a single particle on $N=15$ sites numerically for the exact evolution under the Floquet Hamiltonian Eq. \eqref{eq:Heff}, the fermionic Hamiltonian with modulated hopping energies Eq. \eqref{eq:HFloquet} and the fermionic Hamiltonian approximated by the domain wall Hamiltonian $H_{DW}$ with modulated transverse fields $t_n$. The particle is initially localized in the center of the chain. The resulting site occupation probabilities, as well as the fidelities, are depicted in Figure \ref{fig:FloquetOccupation} (a). We find that the occupation probabilities are well reproduced by the driven domain wall approximation. The fidelity with the exact evolution according to $H_{NNN}$ is above 0.95. The error is mainly caused by the domain wall approximation, indicated by the fact that the evolution under the modulated $H_{NN}$ reaches a fidelity $>0.99$. Note that the asymmetric occupation distribution is a result of asymmetric propagation, caused by the imaginary hopping terms and the breaking of time-reversal symmetry \cite{Novo_2021}.\\
We use a strong ferromagnetic domain wall coupling of $J = -15$. This is necessary for this approach, since now the energy scale of the fermionic system is not only determined by the hopping energies $|t_n^{(0)}| + |t_n^{(1)}|$, but also by the drive frequency $\Omega$, which could excite the system. Therefore, a strong domain wall coupling is required to keep the $M$-subspaces well separated. Alternatively, one could re-scale the system parameters and the simulation time to achieve the same accuracy, as discussed previously.\\
We extend the numerical analysis by performing the same simulation, but with initial states with multiple particles. We compute the fidelity of the simulation as a function of the stroboscopic evolution time $T_{\text{evol}} = m\tau$ with $m \in \mathbb{N}$, when the time evolution of the driven system coincides with the time evolution under the effective Hamiltonian. The result is shown in Figure \ref{fig:FloquetOccupation} (b). We find that while the fidelity $\mathcal{F}$ of the simulation for $M=1$ is remarkably high throughout the simulation, for $M>1$, $\mathcal{F}$ decays substantially faster, but remains reasonably large over the simulated time interval, reaching $\approx 0.74$ at worst.\\
In this section we have shown that by designing a particular periodic drive of the transverse field, exotic terms can be effectively realized in the Hamiltonian. We have used Floquet engineering to create effective imaginary next-nearest-neighbor hopping that the Hamiltonian does not support natively. Let us point out that we have demonstrated a particular application of Floquet engineering in conjunction with the domain wall approximation. There are further intriguing systems that could be investigated in future research, such as time crystals \cite{Estarellas_2020} and effective two-dimensional models \cite{Zou_2017, Lü_2019, Olin_2023}.
\begin{figure*}
    \centering
    \includegraphics[scale=.6]{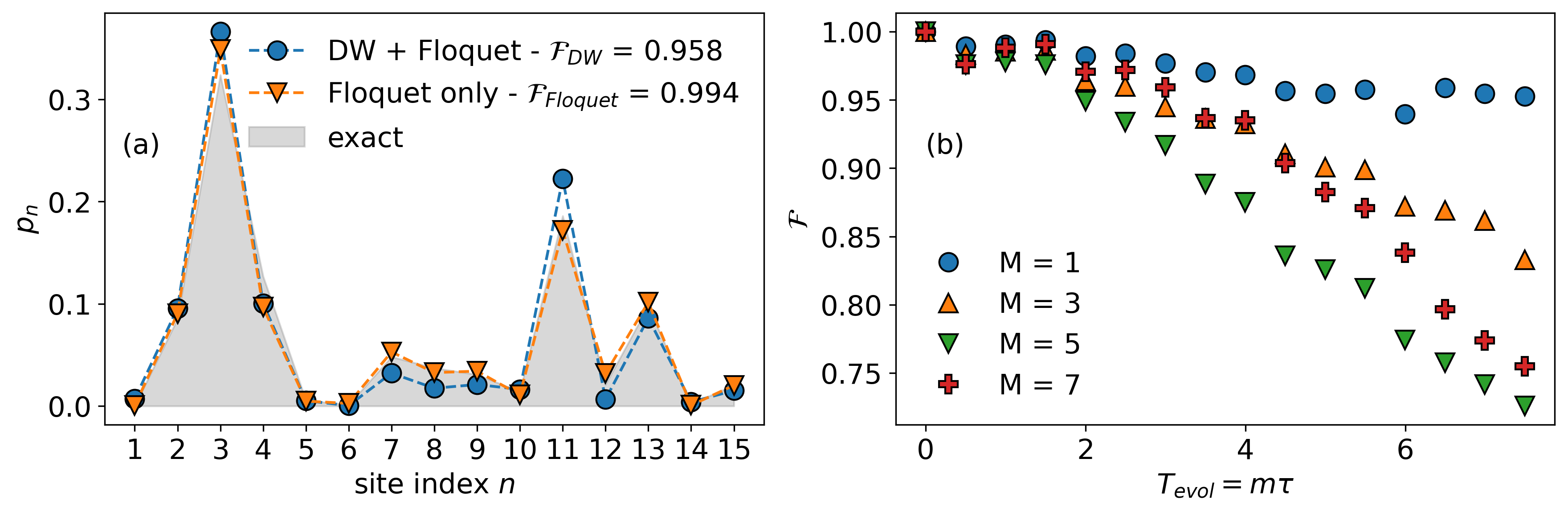}
    \caption{\textbf{(a):} Site occupation $p_n$ after a single particle initially localized at site 8 is evolved under the exact, effective Hamiltonian $H_{NNN}$ Eq. \eqref{eq:Heff} with imaginary next-nearest-neighbor coupling (shaded gray, exact), the fermionic Hamiltonian $H_{NN}$ Eq. \eqref{eq:HFloquet} with periodically driven hopping energies $t_n$ (orange, Floquet only) and the domain wall approximation $H_{DW}$ with periodically driven transverse field (blue, DW+Floquet). $\mathcal{F}_{\text{Floquet}}$ and $\mathcal{F}_{DW}$ are the respective fidelities of the approximately evolved states with the exact state. The DW+Floquet, Floquet only and exact evolution correspond to evolutions under the Hamiltonians Figure \ref{fig:FloquetSchema} (a), (b) and (c) respectively with the appropriate driving. The model parameters are $K_1 = 1$, $K_2 = 0.2$, $\tau = 0.5$ and $T_{\text{evol}} = 15\tau$, while the domain wall coupling is $J=-15$. \textbf{(b):} State fidelity $\mathcal{F}$ as a function of stroboscopic evolution time $T_{\text{evol}} = m \tau$ of the driven domain wall simulation of the effective $H_{NNN}$ for the same model, but with different initial states. The results of $M=1$ correspond to the simulation in (a), while the other initial states in Fock space are $|001000010000100 \rangle$ ($M=3$), $|010010010010010 \rangle$ ($M=5$) and $|010101010101010 \rangle$ ($M=7$) respectively.}
    \label{fig:FloquetOccupation}
\end{figure*}

\section{Conclusion and outlook}
In this work, we propose a method to simulate a variety of spinless fermionic systems in one dimension using an Ising chain with nearest-neighbor ZZ-interactions and transverse X-field to simulate non-interacting particles hopping on a one-dimensional chain. Using next-nearest-neighbor ZZ-interactions, we can control the interactions between particles. We achieve this by treating domain walls on the Ising chain as spinless fermions.\\
The simplicity of the required Hamiltonian is of particular interest, since analog quantum simulators in the noisy intermediate-scale quantum (NISQ)-era will have limited connectivity and limited interaction types. For example, the couplings required to implement hopping terms on platforms such as superconducting flux qubits are highly non-trivial to realize and a matter of active research. Our method makes a large number of canonical quantum many-body systems tractable for simple devices, such as a linear chain of superconducting flux qubits with X-control loops and inductive couplings.\\
The models we have investigated range from the SSH model as a canonical example of a topological insulator, over the Aubry-Andre model displaying a rich phase diagram with localized, critical and extended phases to the inhomogeneous XXZ model with integrable and chaotic phases. Lastly, we show how Floquet engineering can be used to realize imaginary next-nearest-neighbor hopping terms, which gives rise to dynamics with broken time-reversal symmetry. These are systems that display highly non-trivial physics and are only some examples of the Hamiltonians that can be simulated accurately with Ising chains using our domain wall method.\\
A perturbative analysis of the error due to the approximation made in our method showed that the main source of error are perturbations of the energy eigenvalues of the $M$-particle domain wall Hamiltonian $H_{DW}^{(M)}$ leading to phase deviations. These perturbations are caused by interactions with states in subspaces of distinct $M$. The analysis shows that the error can be made arbitrarily small by increasing the strength of the domain wall coupling $|J|$, or alternatively by re-scaling the model parameters at moderate a time-overhead. The re-scaling of the model parameters leads to a time complexity of our continuous time simulation method that scales the same in terms of time and error as first-order product formulas for gate-based quantum simulators.\\
In order to gauge the viability of an implementation of our approach in real quantum hardware, we investigate the impact of decoherence on the simulation fidelity and find that for realistic estimates of the single-qubit relaxation and dephasing error rates, dynamic properties of the Aubry-Andre model can be observed for short time evolutions. This makes our approach a reasonable benchmarking task to compare analog NISQ-devices. Our method allows to reproduce various recent results for the aforementioned models using only a simple Ising Hamiltonian. This moves experimental verification of those results on programmable quantum hardware within reach and will be examined in more detail in the future.\\
Our work suggests various future lines of research, most notably towards the experimental realization of the systems we have discussed here, e.g. using superconducting qubits or Rydberg atoms. Beyond that, we believe it will be insightful to further investigate the possibilities of Floquet engineering in order to realize longer ranging interactions, as well as higher dimensional effective models. Furthermore, while we have investigated both the SSH model, as well as an interacting XXZ chain, our approach also allows to combine the SSH model with nearest-neighbor interactions. This model has been shown to exhibit a rich phase diagram \cite{Yu_2020}. Additionally, the ability of simple Ising Hamiltonians to simulate spin chains makes them a viable candidate as a component in larger quantum information processing architectures. Thus, in the future we intend to examine the quantum state transfer properties of the approach more closely.\\
Furthermore, as we have argued in section \ref{sec:ErrorAnalysis} and verified numerically in section \ref{sec:AAModel}, the error of the simulation is dependent on the initial state and the phase of the system, in that states close to eigenstates of the Hamiltonian are less affected by errors. This result seems closely related to the findings of \cite{Heyl_2019}, where it was shown that errors in localized systems are controllable. It will be fruitful to develop a deeper understanding of these results in relation to our proposed analog quantum simulation method.

\section*{Acknowledgements}
We thank Arnau Riera, Maciej Lewenstein, Marcin Plodzien, Piotr Sierant and Yuma Watanabe for valuable discussions. AGS and MPE acknowledge support by the European Commission FET-Open project AVaQus (GA 899561). MW acknowledges support by the European Commission EIC-Transition project RoCCQeT (GA 101112839) and the Agencia de Gesti\'o d’Ajuts Universitaris i de Recerca through the DI grant (No. DI75).

\appendix
\section{Symmetries and consistent simulations}
\label{sec:AppendixA}
The the main text, we alluded to a relevant point. The particles are domain walls and a domain wall can be "down-up" or "up-down", depending on if on the left of the wall all spins are up and on the right side all spins are down, or vice versa. Since we treat both types of domain walls as identical, this corresponds to a $\mathbb{Z}_2$-symmetry of the system. Any observable, and in fact any operator, acting on the domain walls needs to be invariant under this symmetry in order to be consistent with the fermion interpretation of the domain walls. The Hamiltonian $H_{DW}$ with periodic boundary conditions is indeed an invariant of the spin-flip operator and so is $H_{DW}$ with closed boundary conditions, provided that one also flips the virtual spins. The flipping of the virtual spins corresponds to a change of sign of the local Z-fields, as discussed above. Consequently, the time-evolution under $H_{DW}$ also commutes with the spin-flip symmetry.\\
One consequence of this symmetry consideration is that for periodic boundary conditions, the Hilbert space is twice as large as it should be and there is no one-to-one mapping of fermionic to domain wall states. As an example, the fermionic Fock state $|00110\rangle$ can be represented each by the the two domain wall states $|0010\rangle$ and $|1101\rangle = \Pi |0010\rangle$, where $\Pi = \prod_n \sigma_n^x$ is the spin-inversion operator. Since the Fock states form a basis of the fermionic Hilbert space, any fermion state $|\psi_F\rangle$ can be mapped to two domain wall states that we shall denote as $|\psi_{DW,1}\rangle$ and $|\psi_{DW,2}\rangle = \Pi |\psi_{DW,1}\rangle$ respectively. Note that $|\psi_{DW,1}\rangle$ and $|\psi_{DW,2}\rangle$ are related via the spin-inversion symmetry.\\
The Hamiltonian $H_{DW}$, as well as any other operator $O$ that is consistent with the domain wall encoding, commutes with $\Pi$. Therefore, the Hilbert space is now divided into two subspaces of equal dimension, spanned by eigenstates of $\Pi$ with eigenvalue $\pm 1$ respectively. Consistent operators act the same upon each subspace. In fact, $H_{DW}$ is equivalent to $H_{\text{Fermi}}$ in both subspaces, so one way to mitigate the non-uniqueness of the mapping from fermion states to domain wall states could be to restrict to either of the $\pm1$-eigenspaces of $\Pi$ and the fermionic state could be mapped to domain wall states according to
\begin{equation}
    |\psi_F \rangle \rightarrow \frac{1}{\sqrt{2}} (|\psi_{DW, 1} \rangle \pm |\psi_{DW, 2} \rangle) \ .
\end{equation}
However, this would require the preparation of highly entangled initial states to run the quantum simulation, as can be seen by the example above, where the mapping reads
\begin{equation}
    |00110\rangle \rightarrow \frac{1}{\sqrt{2}} (|0010\rangle \pm |1101\rangle) \ .
\end{equation}
The state on the right-hand side is a GHZ state, up to a flip of the third bit.\\
Luckily, it can be shown that one does not require the restriction to one of the eigenspaces of $\Pi$. Instead, it is consistent to use a more general mapping according to
\begin{equation}
    |\psi_F \rangle \rightarrow |\psi_{DW} \rangle = \alpha|\psi_{DW, 1} \rangle \pm \beta|\psi_{DW, 2} \rangle
    \label{eq:ConsitentState}
\end{equation}
where $|\alpha|^2 + |\beta|^2 = 1$. Consider now the expectation value of a consistent observable $O$ measured on $|\psi_{DW} \rangle$
\begin{equation}
    \begin{aligned}
        &\langle \psi_{DW} | O |\psi_{DW} \rangle \\
        = &|\alpha|^2 \langle \psi_{DW,1} | O | \psi_{DW,1} \rangle + |\beta|^2 \langle \psi_{DW,2} | O | \psi_{DW,2} \rangle \\
        = &|\alpha|^2 \langle \psi_{DW,1} | O | \psi_{DW,1} \rangle + |\beta|^2 \langle \psi_{DW,1} | O | \psi_{DW,1} \rangle \\
        = &\langle \psi_{DW,1} | O | \psi_{DW,1} \rangle
    \end{aligned}
\end{equation}
where in the second line we use $\Pi O \Pi^\dagger = O$, $\Pi=\Pi^\dagger$ and $|\psi_{DW, 1} \rangle = \Pi |\psi_{DW,2}\rangle$. Similarly, one can show that
\begin{equation}
    \langle \psi_{DW} | O |\psi_{DW} \rangle = \langle \psi_{DW,2} | O | \psi_{DW,2} \rangle = \langle \psi_{DW,1} | O | \psi_{DW,1} \rangle \ .
\end{equation}
Therefore, all expectation values of consistent operators are identical, no matter if they are measured in the $\pm1$-eigenspaces of $\Pi$ or in a superposition state, as long as the state is only a superposition of symmetrically related states, as shown in Eq. \eqref{eq:ConsitentState}. In particular, one could choose $\alpha = 1$, $\beta = 0$ and initialize the system only in one symmetry sector. For the example of the Fock state $|00110\rangle$, this means that it is also allowed to initialize the simulation in either of the product states $|0010\rangle$ or $|1101\rangle$. Note, that $O$ can also be a time-evolved operator $O_T = U_T^\dagger O U_T$, where $U_T$ is the time-evolution under a consistent Hamiltonian. Hence, it is not necessary to prepare the system initially in a highly entangled state in order to prepare an initial Fock state. Instead, it is also possible to prepare one of the corresponding domain wall states.

\section{Mapping operators to domain wall encoding}
\label{sec:AppendixB}
In appendix \ref{sec:AppendixA} we discussed the conditions under which we can trust the results of the simulation. To this end, we introduced the concept of consistent operators, which are simply operators that are invariant under the symmetry of the domain wall encoding. In this section we show how spin- and fermionic operators can be mapped to the domain wall picture.\\
To start, note that any Hermitian operator $O$ in the $2^N$-dimensional Hilbert space can be expressed in Pauli-strings $P_\sigma$
\begin{equation}
    \begin{aligned}
        O &= \sum_{P_\sigma \in \{ \mathbb{I}, \sigma^x, \sigma^y, \sigma^z \}^N} o_{\sigma} P_\sigma \ .
    \end{aligned}
    \label{eq:OpDecomposition}
\end{equation}
Using the identities
\begin{equation}
    \begin{aligned}
        \sigma^x &= \frac{1}{2} (\sigma^+ + \sigma^-) \\
        \sigma^y &= \frac{1}{2i} (\sigma^+ - \sigma^-) \\
        \sigma^z &= 2 \sigma^+ \sigma^- - 1
    \end{aligned}
\end{equation}
the operator $O$ can equivalently be decomposed into strings of $\sigma^+$ and $\sigma^-$.\\
An important observation is that the parity of the particle number in the domain wall picture is preserved. This is easy to see, as depending on the construction of $H_{DW}$ either even or odd $M$ can be realized exclusively. Consequently, after projecting into the respective subspace of particle number parity, only Pauli strings with an even number of $\sigma^+$ and $\sigma^-$ can contribute to the decomposition of $O$. Hence, the fundamental building blocks of operators will not be $\sigma^-_i$ and $\sigma^+_i$ acting on site $i$, but $\sigma_i^\pm \sigma_j^\pm$ and $\sigma_i^\mp \sigma_j^\pm$ acting on (possibly identical) sites $i$ and $j$.\\
Let us assume $i \leq j$. By flipping all spins of the Ising chain between $i-1$ and $j$, we can create or destroy domain walls on site $i$ and $j$. The flipping of the spins between $i-1$ and $j$ is captured by the operator
\begin{equation}
    X_{ij} := \prod_{n = \min(i,j)}^{\max(i,j)-1} \sigma_n^x \ .
\end{equation}
The minimum and maximum in the indices generalize to cases where $i > j$. As a convention, if the index $n$ does not run over positive numbers, $X_{ij} = \mathbb{I}$, which is particularly the case if $i=j$. Whether a domain wall is created or destroyed depends on there already being domain walls on the respective site. Consequently, the operator $X_{ij}$ acting on the Ising chain can be identified with the spin chain operator
\begin{equation}
    X_{ij} \rightarrow \sigma_i^+ \sigma_j^+ + \sigma_i^+ \sigma_j^- +\sigma_i^- \sigma_j^+ +\sigma_i^- \sigma_j^- \ .
    \label{eq:FlipOp}
\end{equation}
For domain walls, we can identify the number operators
\begin{equation}
    \begin{aligned}
        n_i &= \frac{1}{2} \left(\mathbb{I} -\sigma_{i-1}^z \sigma_{i}^z  \right) \rightarrow \sigma^+_i \sigma^-_i \\
        \bar{n}_i &= \frac{1}{2} \left(\mathbb{I} +\sigma_{i-1}^z \sigma_{i}^z  \right) \rightarrow \mathbb{I} - \sigma^+_i \sigma^-_i = \sigma^-_i \sigma^+_i \ .
    \end{aligned}
    \label{eq:NumberOp}
\end{equation}
This definition is valid in case of ferromagnetic domain wall coupling $J < 0$. For anti-ferromagnetic coupling, the sign of the ZZ-term needs to be flipped. Using Eq. \eqref{eq:FlipOp} and Eq. \eqref{eq:NumberOp}, one can show that
\begin{equation}
    \begin{aligned}
        n_i X_{ij} n_j &\rightarrow \sigma_i^+ \sigma_j^- \ \ \ \ \ \ 
        n_i X_{ij} \bar{n}_j \rightarrow \sigma_i^+ \sigma_j^+ \\
        \bar{n}_i X_{ij} n_j &\rightarrow \sigma_i^- \sigma_j^- \ \ \ \ \ \ 
        \bar{n}_i X_{ij} \bar{n}_j \rightarrow \sigma_i^- \sigma_j^+ \ .
    \end{aligned}
    \label{eq:OpMapping}
\end{equation}
We point out that this mapping has a certain resemblance with the Jordan-Wigner transformation, where local operators also get mapped to non-local ones. Note that in the case of closed boundary conditions, the spins $i=0$ and $i=N$ are virtual spins fixed in spin-up and spin-down respectively. The corresponding $\sigma^z$ in the operators are consequently replaced by $\pm 1$. For the example of the occupation number operators, this means
\begin{equation}
    \begin{aligned}
        n_1 &\rightarrow \frac{1}{2} \left(\mathbb{I} - \sigma_{1}^z  \right) \\
        n_N &\rightarrow \frac{1}{2} \left(\mathbb{I} + \sigma_{N-1}^z  \right) \ .
    \end{aligned}
\end{equation}
Using the mapping Eq. \eqref{eq:OpMapping} in conjunction with the Jordan-Wigner transformation, any Hermitian operator acting on spins or fermions can be expressed as an operator acting on domain walls, with the restriction that the decomposition Eq. \eqref{eq:OpDecomposition} only contains operator strings of even length. While this seems like a substantive constraint at first glance, there are two things to consider for context. First, this is in line with the parity superselection rule for fermionic systems \cite{Vidal_2021} and, second, our approach effectively models a class of fermionic Hamiltonian that preserves the particle number. Naturally, this preserves also the particle number parity and expectation values implicitly project any operator in the respective parity subspace. Therefore, any relevant physical observable can be mapped to the domain wall picture.\\
Let us highlight that, in general, operators $O$ mapped into the domain wall picture can be highly non-local. However, the occupation number operators $n_i$ are of particular interest for many experiments and they are at most two-local. Measuring the occupation of a site merely requires measuring of the ZZ-correlations between neighboring sites in the Ising chain.

\section{Relationship between $\varepsilon_{\| \cdot \|}$ and $\varepsilon$}
\label{sec:AppendixC}
As discussed in the main text, we computed the error in terms of the infidelity of the simulation, which we denote as $\varepsilon$. The error used in the gate complexity of discretized simulations of continuous time evolutions is formulated in terms of the spectral norm of the distance between the respective unitaries \cite{Childs_2019}, which we denote as $\varepsilon_{\| \cdot \|}$. In order to compare the error scaling between Trotterization and our analog approach, we need to compare these two error metrics.\\
In the main text, we denote the target unitary as $U_T$, while its approximation is denoted with $\Tilde{U}_T$, which we can decompose into the target unitary and an error as
\begin{equation}
    \Tilde{U}_T = U_\varepsilon U_T
\end{equation}
By choosing an appropriate basis, we can compare the error measured in terms of the fidelity and the distance without loss of generality by considering the identity operator $\mathbb{I}$ and the error unitary $U_\varepsilon = \sum_n e^{-i\phi_n} |\phi_n \rangle \langle \phi_n|$. Furthermore, since we are interested in the asymptotic behaviour for small errors, we will assume the arguments of the eigenvalues of $U_\varepsilon$ to be small, i.e. $|\phi_n| \ll 1$. We first calculate the spectral norm of the distance
\begin{equation}
    \begin{aligned}
        \varepsilon_{\| \cdot \|} &= \| \mathbb{I} - U_\varepsilon \| \\
        &= \sqrt{\lambda_{\max} ((\mathbb{I} - U_\varepsilon)^\dagger (\mathbb{I} - U_\varepsilon))} \\
        &= \sqrt{\lambda_{\max} (2\mathbb{I} - U^\dagger_\varepsilon - U_\varepsilon)} \\
        &= \sqrt{\max_n \{ 2-2\cos \phi_n \} } \\
        &\approx |\phi_{\max}|
    \end{aligned}
    \label{eq:NormError}
\end{equation}
where $\lambda_{\max}(O)$ denotes the maximum eigenvalue of the operator $O$  and $\phi_{\max} := \max_n \phi_n$. Equivalently, we can compute the fidelity
\begin{equation}
    \begin{aligned}
        \mathcal{F} &= |\langle \Psi_0 |U_T^\dagger \Tilde{U}_T | \Psi_0 \rangle |^2 = |\langle \Psi_T | U_\varepsilon | \Psi_T \rangle |^2 \\
        &= \left| \sum_n |\Tilde{a}_n|^2 e^{-i\phi_n} \right|^2 \\
        &= \sum_{m,n} |\Tilde{a}_n|^2 |\Tilde{a}_m|^2 e^{i(\phi_m - \phi_n)} \\
        &= \sum_n |\Tilde{a}_n|^4 + 2 \sum_{m>n} |\Tilde{a}_m|^2 |\Tilde{a}_n|^2 \cos(\phi_m - \phi_n)
    \end{aligned}
\end{equation}
where we write $U_T |\Psi_0 \rangle = |\Psi_T \rangle = \sum_n \Tilde{a}_n |\phi_n \rangle$ in the eigenbasis of the error $U_\varepsilon$. As before, we consider small errors and the above expression can be approximated as
\begin{equation}
    \begin{aligned}
        \mathcal{F} &\approx \sum_n |\Tilde{a}_n|^4 \\
        &+ 2 \sum_{m>n} |\Tilde{a}_m|^2 |\Tilde{a}_n|^2 \left( 1- \frac{1}{2} ( \phi_m - \phi_n )^2 \right) \\
        &= 1 - \sum_{n, m>n} |\Tilde{a}_m|^2 |\Tilde{a}_n|^2 ( \phi_m - \phi_n )^2
    \end{aligned}
\end{equation}
and subsequently lower bounded by
\begin{equation}
    \begin{aligned}
        \mathcal{F} &\geq 1 - 4 \phi_{\max}^2 \sum_{n, m>n} |\Tilde{a}_m|^2 |\Tilde{a}_n|^2 \\
        &= 1 - 2 \phi_{\max}^2 \left(1 - \sum_{n} |\Tilde{a}_n|^4 \right) \ .
    \end{aligned}
    \label{eq:Infidelity}
\end{equation}
The lower bound on the fidelity corresponds to the maximum of the eigenvalues considered in the operator norm in Eq. \eqref{eq:NormError}.\\
From Eq. \eqref{eq:NormError}, Eq. \eqref{eq:Infidelity} and with the definition of the infidelity as $\epsilon = 1-\mathcal{F}$, we find that $\varepsilon \leq O(\phi_{\max}^2) = O \left( \varepsilon_{\| \cdot \|}^2 \right)$. As discussed in the main text, this shows that our continuous time quantum simulation algorithm has the same error scaling as first-order Trotterization.\\
Note that in the literature the fidelity of a unitary is often given in terms of the operator fidelity $\mathcal{F}_{op} = \frac{1}{d^2} |\Tr(U_\varepsilon) |^2$, which is a lower bound to the fidelity we used here. This can be seen by considering that
\begin{equation}
    \begin{aligned}
        \mathcal{F}_{op} &= \frac{1}{d^2} |\Tr(U_\varepsilon) |^2 = \frac{1}{d^2} \left( \sum_n e^{i\phi_n} \right) \left( \sum_n e^{-i\phi_n} \right) \\
        &= \frac{1}{d^2} \left( d  + \sum_{m>n} 2\cos (\phi_n - \phi_m )\right) \ .
    \end{aligned}
\end{equation}
As before, this can be bounded using the maximum $\phi_{\max}$ and subsequently expanded for small arguments as
\begin{equation}
    \begin{aligned}
        \mathcal{F}_{op} &\geq \frac{1}{d^2} \left( d + (d^2 -d) \cos 2 \phi_{\max} \right) \\
        &\approx \frac{1}{d} + \left( 1 - \frac{1}{d} \right) (1 - 2\phi_{\max}^2) \\
        &= 1 - 2\phi_{\max}^2 \left( 1 - \frac{1}{d} \right) \ .
    \end{aligned}
\end{equation}
where in the approximation we use the Taylor expansion of the cosine. Hence, for small errors, the infidelity using this alternative definition is also $\varepsilon \leq O(\phi_{\max}^2)$.\\
The sum $\sum_n |\Tilde{a}_n|^4$ can be bounded according to
\begin{equation}
    \frac{1}{d} \leq \sum_n |\Tilde{a}_n|^4 \leq 1 \ ,
\end{equation}
which can easily be shown using the Cauchy-Schwarz inequality. Hence, the fidelity of a specific initial state is lower bounded by the operator fidelity. Furthermore, the sum $\sum_n |\Tilde{a}_n|^4$ is related to the localization of the evolved state in the eigenbasis of the error via the participation ratio. In the main text we have argued that the leading error is caused by perturbations on the energy eigenvalues. This effectively translates to an error $U_\varepsilon$ that is diagonal in the eigenbasis of the target Hamiltonian. Thus, the calculation here provides an alternative argument why extended states in the eigenbasis of the time-evolution operator are more affected by errors than localized states.

\bibliographystyle{unsrt}
\bibliography{references}

\end{document}